\documentclass[acmtog, screen, nonacm]{acmart}

\AtBeginDocument{
  }

\usepackage{xcolor}
\usepackage{bm}
\usepackage{algorithm}
\usepackage{algorithmic}
\usepackage{multicol}
\usepackage{multirow}
\usepackage{wrapfig}
\usepackage{subcaption}
\usepackage{colortbl}

\newcommand{\bx}{\mathbf{x}}

\newcommand{\bA}{\mathbf{A}}

\newcommand{\bD}{\mathbf{D}}
\newcommand{\bE}{\mathbf{E}}

\newcommand{\bH}{\mathbf{H}}

\newcommand{\bK}{\mathbf{K}}
\newcommand{\bL}{\mathbf{L}}
\newcommand{\bM}{\mathbf{M}}

\newcommand{\bP}{\mathbf{P}}

\newcommand{\bR}{\mathbf{R}}

\setcopyright{none}
\renewcommand\footnotetextcopyrightpermission[1]{}

\begin{document}

\title{ParaBMG: A Parallel Block Multigrid Framework for Large-Scale Evolving Sparse Systems on Unstructured Meshes}

\author{Kangbo Lyu}
\email{lkb23@mails.tsinghua.edu.cn}
\affiliation{%
  \institution{Tsinghua University and Shanghai Qi Zhi Institute}
  \country{China}
}

\author{Mingqi Geng}
\email{mg2248@cam.ac.uk}
\affiliation{%
  \institution{University of Cambridge}
  \country{UK}
}

\author{Jinchen Jiang}
\email{jiangjc22@mails.tsinghua.edu.cn}
\affiliation{%
  \institution{Tsinghua University and Shanghai Qi Zhi Institute}
  \country{China}
}

\author{Yu Wang}
\email{wangyu9@csail.mit.edu}
\affiliation{%
  \institution{Massachusetts Institute of Technology}
  \country{USA}
}

\author{Minchen Li}
\email{minchernl@gmail.com}
\affiliation{%
  \institution{Carnegie Mellon University and Genesis AI}
  \country{USA}
}

\author{Tao Du}
\correspondingauthor
\email{taodu@tsinghua.edu.cn}
\affiliation{%
  \institution{Tsinghua University  and Shanghai Qi Zhi Institute}
  \country{China}
}

\begin{abstract}
  Solving large-scale, evolving sparse linear systems on unstructured meshes remains the primary performance bottleneck in many GPU-accelerated geometry processing pipelines. Although modern linear solvers like multigrid and multilevel additive Schwarz (MAS) methods offer fast convergence and strong parallelization potential, a unified benchmark and solver design tailored for geometry processing tasks have been lacking. Addressing this gap, we present ParaBMG, a parallel block multigrid preconditioning framework that combines fully parallel geometric hierarchy construction, fast block-Jacobi smoothing, and GPU-optimized operators. We develop CUDA implementations of four standard geometry processing pipelines to form a large-scale benchmark for GPU linear solvers, comprising more than 40K linear systems with up to 6.6M unknowns. Across comparisons in this dataset, ParaBMG achieves speedups of 3.7x-7.7x over strong general-purpose algebraic multigrid baselines,  6x-105x speedups over our custom GPU implementations of the state-of-the-art CPU multigrid solvers for Poisson problems, and 6x-96x  speedups over MAS. Compared with open-sourced multigrid baselines, ParaBMG reduces repeated hierarchy construction costs  through hierarchy reuse and improves convergence through MAS-inspired block-Jacobi smoothing. Compared with MAS, the complementary coarse corrections reduce iteration counts by an order of magnitude while maintaining low per-iteration costs. It enables three interactive geometry processing applications with order-of-magnitude speedups over their public implementations.
\end{abstract}


\begin{teaserfigure}
\includegraphics[width=\textwidth]{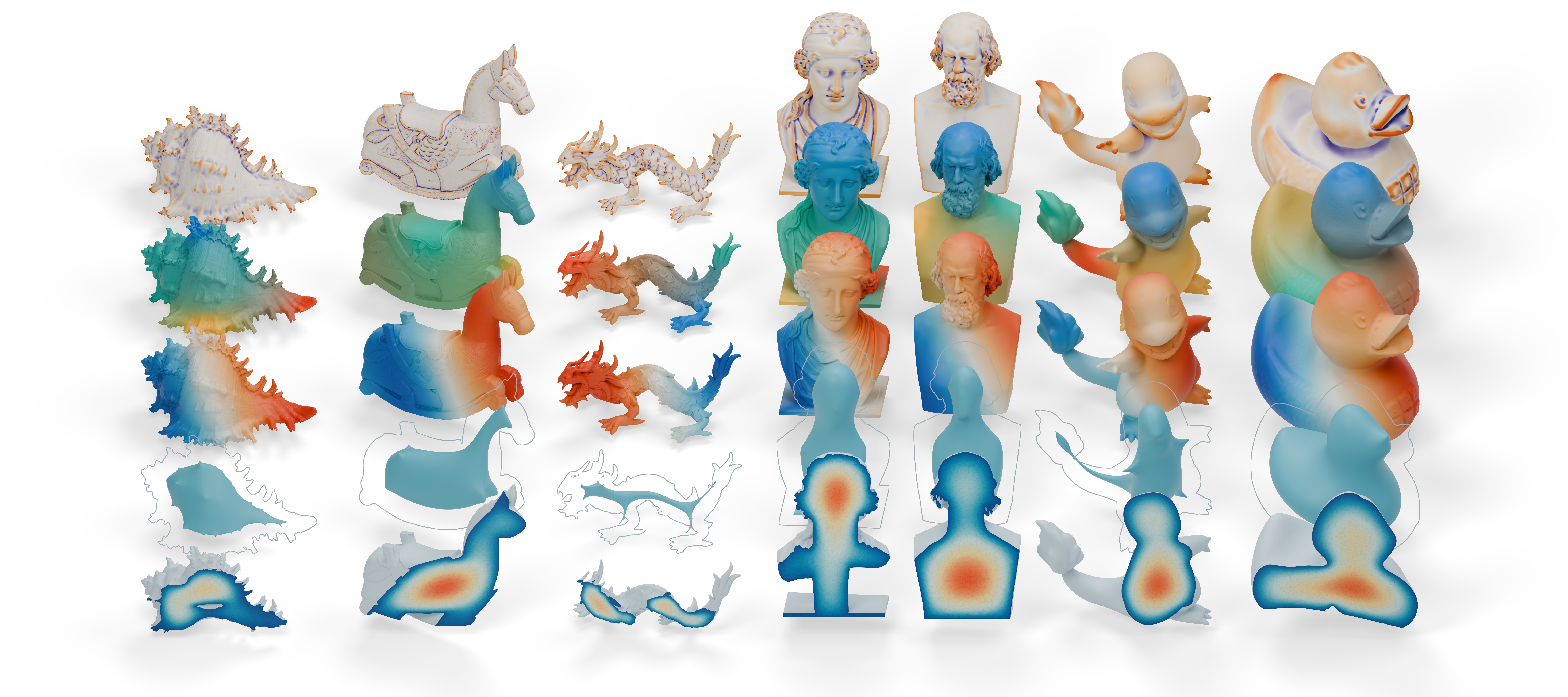}
\caption{Our ParaBMG enables multiple geometry processing problems for large-scale meshes. From top to bottom, the row shows directional shape total variation (DSTV), quasi-harmonic   weights (QHW), screened surface Poisson, inverse mean-curvature-flow (IMCF) optimization, and volumetric Poisson. DSTV is visualized through retained signed spectral components on the input surface. QHW is demonstrated by surface weight scales. IMCF shows MCF-contracted shapes before inverse optimization, with faint outlines indicating the final geometries. The volumetric Poisson solutions are revealed through tetrahedral cross-sections.}
\Description{This is a description}
\end{teaserfigure}

\maketitle

\section{Introduction}
Geometry processing pipelines, such as mesh editing,
surface parameterization, and mean curvature flow-based inflation, often require
repeatedly solving large-scale sparse linear systems.
Their matrix coefficients evolve as the geometry or optimization
variables change across optimization iterations or time steps.
As the mesh resolution increases, these
solves often dominate the overall computation and limit
interactive performance.
High-performance solvers for evolving linear systems are
therefore critical to large-scale geometry processing.

Multilevel solvers accelerate convergence by combining local corrections with coarse-scale corrections to address error components across spatial scales \cite{briggs2000multigrid}. Multiplicative formulations, including the multigrid (MG) framework adopted in this work, apply these
corrections successively and admit sharper convergence bounds under suitable assumptions
  than their additive counterparts \cite{xu1992iterative} such as the multilevel additive Schwarz (MAS) frameworks \cite{wu2022mas}.
Despite recent progress in high-performance MG and MAS solvers, challenges remain for evolving linear systems in geometry processing. General-purpose algebraic multigrid (AMG)
solvers~\cite{mills2021petsc,henson2002boomer,demidov2019amgcl,naumov2015amgx} provide mature GPU implementations, but  do not exploit
mesh geometry and incur additional update costs as coefficients evolve. Unstructured mesh-based geometric multigrid (GMG) methods~\cite{liu2021surface, wiersma2023gravo,padilla2026gravo}
exploit the geometry structure, yet fully parallel GPU implementations
with effective GPU-friendly smoothers remain underexplored.

Meanwhile, MAS~\cite{wu2022mas,huang2023gipc,huang2024stiffgipc,wang2026agipc}
is the state-of-the-art GPU preconditioner for elasticity simulation,
where it outperforms general-purpose AMG; each implicit simulation step resembles a strongly screened
Poisson-type system. Geometry processing pipelines instead produce weakly screened or
unscreened Poisson-type systems with mesh-wide coupling.
As screening weakens, MAS's iteration penalty relative to multigrid
grows until its lower per-iteration cost no longer compensates,
and the most effective solver shifts from additive multilevel
methods, whose inexpensive local corrections suffice when coupling is
short-range, toward multiplicative multigrid, whose successive
coarse-grid corrections reduce mesh-wide errors more
effectively~\cite{xu1992iterative}.
Solver design for geometry processing therefore cannot simply be
inherited from simulation.

To bridge this gap, we present \textbf{ParaBMG}, a parallel block multigrid framework for evolving sparse systems on unstructured meshes. The MAS with non-overlapping domains can be viewed as block-Jacobi smoothing that enables efficient local error reduction and highly parallel execution, while multigrid coarse spaces are essential for effectively eliminating low-frequency global errors. Inspired by MAS \cite{wu2022mas} and the graph partition-based hierarchy construction in GMG \cite{wiersma2023gravo}, ParaBMG uses the mesh partition at each level both to construct the geometric hierarchy and to configure the block-Jacobi smoother,
and organizes matrix storage around the resulting smoothing block partitions.
Specifically, it stores the diagonal blocks densely and the cross-block entries sparsely, allowing smoothing, residual evaluation, and coarse matrix updates to operate directly on this block-structured storage.
The hierarchy and block partitioning are constructed on the GPU and reused as matrix coefficients evolve.
 Prepared coarse matrix updates, in-place block inversion, and adaptive spectrum estimation further reduce the recurring cost of solving the evolving linear systems.

To benchmark existing solvers and evaluate ParaBMG, we develop
CUDA implementations of four representative geometry processing
pipelines, including local mesh deformation, skinning-weight
computation, feature-preserving mesh filtering, and mesh
inflation through inverse optimization of mean-curvature flow.
The dataset consists of 40,987 linear solves
on 20 triangular or tetrahedral meshes with 177K to 6.6M vertices.
Across the benchmark dataset, ParaBMG achieves speedups of $3.7\times$–$10.7\times$ over AMGX, $4.5\times$–$8.5\times$ over BoomerAMG, $6\times$–$105\times$ over our customized GPU
  implementations of GravoMG and GravoTet, and $6\times$–$96\times$ over a SOTA MAS implementation.  Our ablation studies further reveal that
large smoothing blocks improve MG convergence, while
MG coarse corrections substantially reduce the
iteration counts of MAS.
We further examine the effects of block size, smoother
adaptation, and optimized GPU operators, showing how
convergence gains and execution costs jointly determine
overall performance. We demonstrate our efficacy on three interactive applications for high-resolution mesh processing, including mesh
editing~\cite{chen2023ldef}, surface parameterization with
changing boundary constraints~\cite{wang2023qhm}, and quasi-harmonic
clustering around handles~\cite{wang2021qhw}.
These applications achieve order-of-magnitude speedups
over their SOTA CPU implementations, enabling interactive
processing on large-scale meshes of up to 10.2M vertices that previously
struggled to handle.

Our main contributions are summarized as follows:
\begin{itemize}
    \item  We propose \textbf{ParaBMG}, a multigrid framework for preconditioning that couples a reusable geometric multigrid hierarchy with partition-induced block-Jacobi smoothing in a fully parallel GPU framework for evolving linear systems.

    \item
    We design a GPU execution pipeline for ParaBMG tailored to evolving
systems, integrating fully parallel hierarchy construction
and reuse with in-place coarse matrix updates, efficient block
inversion, and prepared coarsest solves to reduce recurring
costs.

\item
We establish a unified GPU benchmark through CUDA
implementations of four geometry processing pipelines,
covering more than 40K systems on meshes with up to
6.6M vertices. ParaBMG achieves an overall
$3.7\times$-$105\times$ speedup over the strong GPU baselines including
  general-purpose AMG, geometric multigrid, and MAS baselines.

\item
  Our comparisons and ablation studies demonstrate that ParaBMG's
  block-aware hierarchy reduces solver cost, partition-induced
  block-Jacobi smoothing improves convergence, and optimized GPU
  execution reduces the solving time while preserving convergence.

    \item
   We demonstrate that ParaBMG enables three geometry processing
applications to operate interactively on previously impractical
mesh scales, delivering order-of-magnitude 
speedups over their open-source implementations.
\end{itemize}

\section{Related Work}
\subsection{Large-Scale Linear Solvers on Unstructured Meshes}

Large-scale sparse symmetric positive definite (SPD) systems on unstructured meshes can be solved using sparse direct solvers or iterative methods. While direct solvers provide robust solutions, their memory requirements and factorization costs can limit scalability. Preconditioned conjugate gradient (PCG) offers an alternative, whose efficiency depends on both the convergence rate and the cost of preparing and applying the preconditioner. Multigrid (MG) methods accelerate convergence through local smoothing and coarse-grid correction \cite{briggs2000multigrid}, and can serve as standalone solvers or preconditioners within PCG.

General-purpose algebraic multigrid (AMG) frameworks, such as BoomerAMG~\cite{henson2002boomer}, AmgX~\cite{naumov2015amgx}, and AMGCL~\cite{demidov2019amgcl}, employ multigrid cycles as preconditioners within PCG to accelerate convergence on large-scale sparse systems. However, these general-purpose frameworks do not fully exploit fixed sparsity patterns across evolving systems, incurring unnecessary overhead from repeated hierarchy construction and numerical updates.

Problem-specific GPU solvers have also been extensively developed for simulation problems recently, including highly optimized grid-based Poisson solvers \cite{mcadams2010gmg, Shao2022uaamg, sun2025lfm} and multilevel additive Schwarz methods~\cite{wu2022mas,huang2023gipc,huang2024stiffgipc,wang2026agipc} on unstructured meshes. Grid-based solvers, however, do not directly extend to unstructured meshes, while additive Schwarz corrections can be less effective at reducing global errors in the Poisson-like problems arising in geometry processing pipelines, leading to slower convergence and more solver iterations. 

For geometry processing, Laplacian preconditioning \cite{krishnan2013efficient} developed a multilevel preconditioner for discrete Laplacian systems. Surface Multigrid \cite{liu2021surface}, GravoMG \cite{wiersma2023gravo}, and GravoTet \cite{padilla2026gravo} developed geometry-aware multigrid solvers with efficient hierarchy construction for unstructured meshes. While these methods provide effective coarse spaces, their hierarchy designs alone do not address the combined challenges of strong parallel smoothing and efficient numerical updates for evolving systems.

ParaBMG addresses these challenges by integrating partition-induced block-Jacobi smoothing with geometric multigrid coarse-grid correction, achieving efficient local and global error reduction.
\subsection{Implementation of Large-Scale Linear Solvers}
High-performance implementation of large-scale linear solvers requires careful optimization of their computational components. Sparse direct solvers, such as PARDISO \cite{schenk2001pardiso} on CPUs and cuDSS \cite{nvidia_cudss} on GPUs, employ highly optimized factorization algorithms. However, fill-in and repeated numerical factorization can impose substantial memory and computational costs on large evolving systems.

For iterative solvers, GPU-based multilevel additive Schwarz (MAS) preconditioned conjugate gradient methods \cite{wu2022mas,huang2023gipc,huang2024stiffgipc,wang2026agipc} demonstrate the benefits of efficient in-place diagonal block matrix inversion and GPU-oriented data layouts. Their implementations exploit highly parallel local block-based operations to achieve low preconditioning overhead. However, their efficiency gains are insufficient to offset the iteration-count overhead in Poisson-like problems, where MG is more advantageous than MAS.

Multigrid is among the most effective approaches for large-scale sparse linear systems, while efficient GPU execution requires careful engineering across its components, each exhibiting distinct parallelism and memory-access patterns. For example, AmgX \cite{naumov2015amgx} employs parallel coarsening and relaxation strategies, while \citeN{geveler2013towards} demonstrate the importance of strong smoothers and degree-of-freedom ordering for GPU geometric multigrid.

Despite these advances, existing GPU multigrid implementations are not specifically optimized for repeated solves of evolving systems with fixed sparsity patterns.
ParaBMG reduces repeated hierarchy construction overhead through structural reuse and improves convergence with MAS-inspired block-Jacobi smoothing, leveraging efficient local block inversion. We further integrate reusable symbolic structures, in-place Galerkin updates and prepared coarsest solves to minimize the overall cost of repeated solves.
\subsection{Geometry Processing Applications and Benchmarks}
Solving large-scale sparse linear systems is a fundamental task in geometry processing, particularly in variational optimization and PDE-based methods. Examples include local deformation for interactive shape editing (LDSE) \cite{chen2023ldef}, which involves sparse linear solves within an alternating direction method of multipliers (ADMM) optimization framework. Variational quasi-harmonic maps (QHM)~\cite{wang2023qhm} and quasi-harmonic weights (QHW) \cite{wang2021qhw} employ parameterized elliptic PDEs to compute diffeomorphic mappings and skinning weights, respectively. \citeN{dodik2026iskra} demonstrated many inverse-problem solving pipelines, including a mesh inflation through inverse mean-curvature-flow (IMCF) optimization requiring repeated linear solves. \citeN{brokman2024dstv} developed directional shape total-variation (DSTV) processing for feature-preserving shape evolution, motivating efficient numerical treatment of nonlinear geometric flows.

In these applications, sparse linear solves become a significant computational bottleneck when mesh resolution increases. Existing pipelines employ different solver designs and implementations, making it difficult to compare solver performance across applications and identify generalizable optimizations. In particular, systematic GPU benchmarks covering different geometry processing workloads remain limited, hindering the development and evaluation of
reusable high-performance solvers.

To address this, we develop CUDA implementations of these five representative applications, establishing a unified benchmark for GPU linear solvers. This benchmark enables systematic comparisons of convergence, numerical-update costs, and overall solving performance across diverse applications. Together with ParaBMG, these implementations accelerate large-scale geometry processing and enable interactive applications on high-resolution meshes.
\section{Background}

We formally define the \emph{evolving linear systems} considered in this work and briefly introduce the background on preconditioning and multigrid.

\label{sec:background}

\subsection{Evolving Linear Systems}
Geometry processing algorithms often require sequences of linear solves in applications such as mesh deformation, parameterization, and mean curvature flows.  Across these different pipelines, the resulting linear systems frequently involve a spatially varying Poisson or screened Poisson structure after discretization, whose coefficients evolve over optimization iterations or physical steps.

Let $\mathcal M=(V,E,\mathcal T)$ be a mesh with vertex set $V$,
edge set $E$, element set $\mathcal T$, and $N=|V|$ vertices.
We consider 3-dimensional problems, so the vertex positions $\{\mathbf{p}_v\}_{v\in V}$ have dimension $\mathbf{p}_v\in \mathbb{R}^3$.
We use $d=2$ for the triangular surface meshes and $d=3$
for the tetrahedral volume meshes.
Each element $\tau\in\mathcal T$ carries a symmetric
positive-definite diffusion coefficient matrix
$\mathbf C_\tau^{(\nu)}\in\mathbb S_{++}^{d\times d}$.
Using piecewise-linear finite elements, the assembled system
matrix takes the form
\begin{align}
  \mathbf A^{(\nu)}
  =
  \sum_{\tau\in\mathcal T}
  |\tau|\,\mathbf G_\tau^\top
  \mathbf C_\tau^{(\nu)}\mathbf G_\tau
  + \mathbf M^{(\nu)},
  \label{eq:matrix_assembly}
\end{align}
where $\mathbf G_\tau\in\mathbb R^{d\times N}$ is the element
gradient operator, whose columns are the vertex basis-function
gradients on $\tau$, with zero columns for vertices outside $\tau$.
Here $|\tau|$ is the element area or volume, and
$\mathbf M^{(\nu)}\in\mathbb S_+^{N\times N}$ is a nonnegative
diagonal matrix representing lumped mass or screening. As the outer non-linear optimizer or physical state steps forward, the linear system index $\nu=0,1,2,\dots$ increases, with the diffusion coefficient matrices $\mathbf C_\tau^{(\nu)}$ and the diagonal term $\mathbf M^{(\nu)}$ slowly change accordingly.

The right-hand-side vector $\mathbf b^{(\nu)}\in{\mathbb{R}^{N}}$ and the solution $\mathbf x^{(\nu)}\in{\mathbb{R}^{N}}$, depend on the specific problem context and corresponding boundary conditions.
Given $\mathbf A^{(\nu)}$ and $\mathbf b^{(\nu)}$, our solver
solves $\mathbf x^{(\nu)}$ from
\begin{align}
  \mathbf A^{(\nu)}\mathbf x^{(\nu)}
  =
  \mathbf b^{(\nu)},
  \qquad \nu=0,1,2,\ldots .
  \label{eq:evolving_system}
\end{align}
We refer to this sequence as \emph{evolving linear systems}
in a geometry processing pipeline.

\subsection{Preconditioning}
\label{sec:bg_precondition}
To solve \autoref{eq:evolving_system} on large-scale meshes,
many state-of-the-art (SOTA) solvers use the preconditioned
conjugate gradient (PCG) method.
Given $\bA^{(\nu)}$, $\mathbf b^{(\nu)}$, and an initial guess
$\bx^{(\nu)}_0$, PCG iteratively improves the solution
$\bx^{(\nu)}_k$ until the residual norm
$\|\mathbf r^{(\nu)}_k\|_2
= \|\mathbf b^{(\nu)}-\bA^{(\nu)}\bx^{(\nu)}_k\|_2$
is sufficiently small, where $k=0,1,\dots,n_{\rm cg}^{(\nu)}$
is the PCG iteration index and $n_{\rm cg}^{(\nu)}$ is the total
number of iterations required for convergence.
At each iteration, the linear preconditioning operator
$\mathcal P^{(\nu)}$ is applied to the residual as
$\mathcal P^{(\nu)}\mathbf r^{(\nu)}_k$.
For standard PCG, $\mathcal P^{(\nu)}$ must be symmetric,
positive definite, and fixed within each solve.
An efficient preconditioner should approximate the matrix
inverse well at low computational cost.

For a sequence of linear solves, the preconditioning pipeline
typically consists of three stages.
The \emph{setup} stage runs once and prepares reusable structures.
The \emph{update} stage refreshes coefficient-dependent data when
$\mathbf C_\tau^{(\nu)}$ and $\mathbf M^{(\nu)}$ change.
The \emph{apply} stage evaluates $\mathcal P^{(\nu)}\mathbf r^{(\nu)}_k$
at each PCG iteration.
For instance, a factorization-based preconditioner performs
symbolic factorization, numerical factorization, and back
substitution at these three stages, respectively.

Our objective is to minimize the total solution time over the
sequence,
$T_{\rm total} = T_{\rm setup}
+\sum_{\nu}(
T_{\rm update}^{(\nu)}
+n_{\rm cg}^{(\nu)}
(T_{\rm apply}^{(\nu)}+T_{\rm cg}^{(\nu)})
)$,
where $T_{\rm setup}$ and $T_{\rm update}^{(\nu)}$ correspond
to the time spent in the \emph{setup} and \emph{update} stages
respectively,
$T_{\rm apply}^{(\nu)}$ is the cost of the \emph{apply} stage, and $T_{\rm cg}^{(\nu)}$ accounts for the
remaining work per iteration.
Preconditioner design balances the iteration count
$n_{\rm cg}^{(\nu)}$ against the costs
$T_{\rm update}^{(\nu)}$ and $T_{\rm apply}^{(\nu)}$.
For a sufficiently long sequence, $T_{\rm setup}$ can be
amortized.
Therefore, we optimize the time cost  per system $T^{(\nu)}$ defined as
\begin{align}
  T^{(\nu)}
  =
  T_{\rm update}^{(\nu)}
  + n_{\rm cg}^{(\nu)}
  \bigl(T_{\rm apply}^{(\nu)}+T_{\rm cg}^{(\nu)}\bigr).
  \label{eq:bg_sequence_cost}
\end{align}

\subsection{Multigrid}
Multigrid (MG) methods \cite{briggs2000multigrid} are a family of iterative solving methods designed for large-scale problems.
They combine iterative smoothing with corrections computed on a hierarchy of smaller proxy problems. A smoother typically reduces error components in high spatial frequencies, while coarse level corrections target the components that the smoother reduces slowly. Geometric multigrid constructs this hierarchy using geometric discretizations, whereas algebraic multigrid constructs it from the system matrix without requiring explicit coarse meshes.

We describe the hierarchy using levels indexed by $l=0,\dots,L$ from finest to coarsest, with $N_l$ unknowns at level $l$.
In our geometric multigrid setting on unstructured meshes, the hierarchy consists of \emph{levels} of meshes $\mathcal{M}_l=(V_l, E_l,\mathcal T_l)$ with $N_l=|V_l|$. 
For each system
$\bA^{(\nu)}\bx^{(\nu)}=\mathbf b^{(\nu)}$,
we set $\bA_0^{(\nu)}=\bA^{(\nu)}$.
The \emph{prolongation} operator
$\bP_l\in\mathbb R^{N_l\times N_{l+1}}$
maps coarse-level corrections to level $l$, while the
\emph{restriction} operator
$\bR_l\in\mathbb R^{N_{l+1}\times N_l}$
transfers residuals to level $l+1$.
Coarse matrices can be obtained by rediscretizing the underlying
continuous problem on coarse meshes or by Galerkin projection.
In this work, we use $\bR_l=\bP_l^\top$ and construct the coarse-level operators recursively as
\begin{align}
  \bA_{l+1}^{(\nu)}
  = \bR_l\bA_l^{(\nu)}\bP_l,
  \qquad l=0,\dots,L-1.
  \label{eq:bg_galerkin}
\end{align}

A \emph{smoother} at level $l$ performs iterative linear solving. For instance, given the right-hand-side vector  $\mathbf{b}_l\in \mathbb{R}^{N_l}$ at level $l$, the damped Jacobi smoother updates the approximated solution $\mathbf{x}_l\in \mathbb{R}^{N_l}$ as 
\begin{align}
  \mathbf x_l
  \leftarrow \mathbf x_l
  + \omega_l^{(\nu)}
  \operatorname{diag}(\bA_l^{(\nu)})^{-1}
  (\mathbf b_l-\bA_l^{(\nu)}\mathbf x_l),
  \label{eq:bg_jacobi}
\end{align}
where $\omega_l^{(\nu)}\in\mathbb{R}_+$ is the damping weight.
Let $\mathcal S_{l,\rm pre}^{(\nu)}$ and
$\mathcal S_{l,\rm post}^{(\nu)}$ denote pre- and post-smoothing. Both take inputs $\mathbf{x}_l, \mathbf{b}_l$ and update $\mathbf{x}_l$ as output. In practice, the number of smoothing operators applied at each level can be tuned. We use two pre-smoothing iterations and two post-smoothing iterations in this work to obtain a good averaged performance.

The \emph{V-cycle} $\mathcal V_l^{(\nu)}(\mathbf b_l)$
is defined recursively for $l<L$ by
\begin{align}
  \mathbf x_l
  &\leftarrow
  \mathcal S_{l,\rm pre}^{(\nu)}(\mathbf 0,\mathbf b_l),
  \nonumber\\
  \mathbf e_{l+1}
  &\leftarrow
  \mathcal V_{l+1}^{(\nu)}
  \bigl(\bR_l(\mathbf b_l-\bA_l^{(\nu)}\mathbf x_l)\bigr),
  \nonumber\\
  \mathbf x_l
  &\leftarrow
  \mathcal S_{l,\rm post}^{(\nu)}
  (\mathbf x_l+\bP_l\mathbf e_{l+1},\mathbf b_l).
  \label{eq:bg_v_cycle}
\end{align}
The cycle returns the updated $\mathbf x_l$, with
$\mathcal V_L^{(\nu)}(\mathbf b_L)
=(\bA_L^{(\nu)})^{-1}\mathbf b_L$
at the coarsest level.
With suitable symmetric smoothing, the V-cycle defines
an efficient SPD operator that can serve as a PCG
preconditioner.
Its \emph{apply} stage evaluates
$\mathcal P^{(\nu)}\mathbf r_{k}
=\mathcal V_0^{(\nu)}(\mathbf r_{k}^{(\nu)})$. 

 The mesh-based geometric multigrid (GMG) \cite{liu2021surface, wiersma2023gravo, padilla2026gravo} uses mesh geometry to construct reusable levels at the \emph{setup} stage and only renews the coefficients at the \emph{update} stage when solving the evolving linear systems, but its parallel implementation with efficient GPU-friendly smoothers is under-explored. Algebraic multigrid (AMG) \cite{naumov2015amgx, demidov2019amgcl,mills2021petsc, henson2002boomer} generalizes the idea of MGs without relying on meshes, and constructs the levels and transfer operators from the matrix structure and coefficients directly, and therefore accepts general sparse matrix inputs, but its existing implementations are less efficient for evolving linear systems due to the repeated level constructions at each \emph{update} stage.
To reduce the recurring \emph{update} and \emph{apply} costs of MG
for evolving systems, our ParaBMG combines a reusable geometric
hierarchy with GPU-parallel block-Jacobi smoothing and
efficient numerical updates.

\section{Method}
\label{sec:method}

\begin{figure*}[t]
 \centering
 \includegraphics[width=\textwidth]{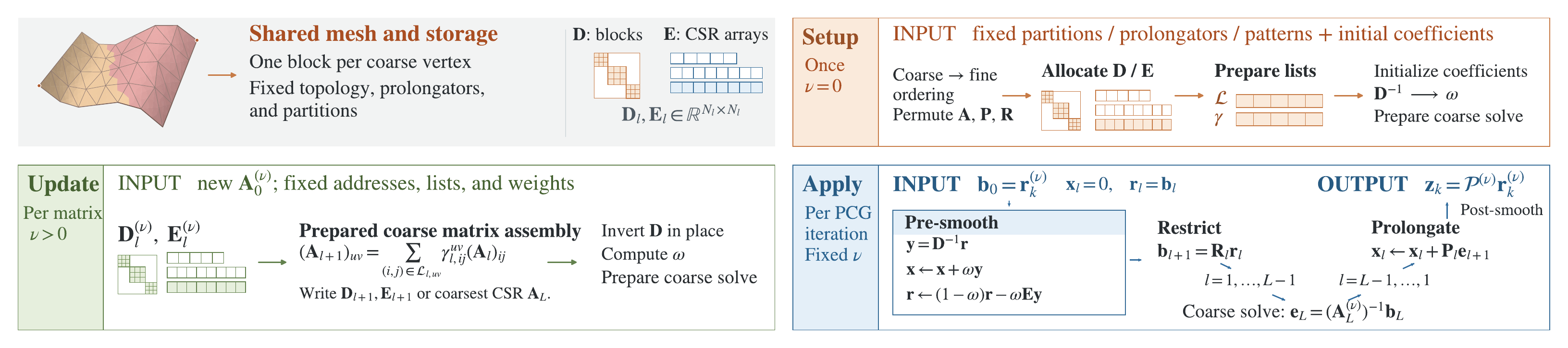}
 \caption{Overview of the ParaBMG pipeline.
 \textbf{Top-right: Setup} constructs the reusable multigrid hierarchy,
 prepares matrix storage and vertex ordering, and initializes the
 numerical data needed for the first solve.
 \textbf{Bottom-left: Update} refreshes coefficient-dependent quantities,
 including coarse matrices, block inverses, smoothing parameters, and
 coarsest-solve operators, while reusing the prepared structure.
 \textbf{Bottom-right: Apply} evaluates a multigrid V-cycle within PCG
 using the fixed prolongation operators and the current numerical data.}
 \Description{The ParaBMG pipeline organized into setup, update,
 and apply stages, showing reusable structures and
 coefficient-dependent numerical data.}
 \label{fig:method_overview}
\end{figure*}

ParaBMG solves evolving linear systems using PCG with a customized
multigrid preconditioner. It combines a geometric hierarchy with
block-Jacobi smoothing, assigning one smoothing block to each coarse
vertex. Following the \emph{setup}, \emph{update}, and \emph{apply}
stages introduced in \autoref{sec:bg_precondition}, ParaBMG separates
reusable structural preparation from coefficient-dependent numerical
updates and repeated preconditioner applications.
\autoref{fig:method_overview} summarizes the resulting pipeline.

We first explain the design rationale behind this organization
(\autoref{sec:method_summary}).
We then describe the parallel construction of the geometric hierarchy,
including block partitions and prolongation operators
(\autoref{sec:method_graph}), followed by the matrix storage and GPU
operators for block-Jacobi smoothing and Galerkin updates
(\autoref{sec:decomposed-block-storage}).
Next, we present power-iteration-based damping estimation for evolving
linear systems (\autoref{sec:method_spectrum}).
Finally, we introduce the customized coarsest-level solver and the
associated vertex ordering
(\autoref{sec:coarse_solve_ordering}).

\subsection{Design Rationale}
\label{sec:method_summary}

ParaBMG  organizes hierarchy construction, block-Jacobi smoothing, and matrix storage around a shared block partition to draw on the complementary strengths of GPU MAS and geometric multigrid and combine effective convergence with efficient GPU execution.
Efficient GPU MAS implementations motivate our use of explicit
block inverses and parallel local solves
\cite{wu2022mas,huang2023gipc,huang2024stiffgipc,wang2026agipc},
while graph-based geometric multigrid methods motivate partitioning
fine-level vertices into blocks associated with coarse vertices
\cite{wiersma2023gravo,padilla2026gravo}.  Together, these
ideas connect efficient local corrections with global error reduction.
We use the same block partition to define the smoothing blocks
and guide the coarse mesh construction. 
This partition also organizes vertex ordering and coefficient
storage, allowing the multigrid components and their parallel
implementation to be designed together.

Our design targets the total cost of solving a sequence of evolving
linear systems.
For systems with fixed geometry and sparsity, the hierarchy,
prolongation operators, block partitions, and coarsest-level symbolic
analysis are prepared once, while coefficient-dependent numerical
data are refreshed for each system. Within this organization, our
customized interpolation balances approximation quality against
update cost, the customized matrix layout efficiently supports both local corrections and cross-block coupling, and the proposed damping adaptation carries spectral information across the evolving systems.
Together, these choices balance convergence, update cost, and
parallel execution efficiency over the full sequence.

\subsection{Reusable Geometric Hierarchy and Block Partitions}
\label{sec:method_graph}
\label{sec:method_details}
\label{sec:coarsening-space-construction}

Repeated solves favor a hierarchy built from the mesh that is independent of the coefficient changing. Starting from $\mathcal M_0=\mathcal M$, we use
mesh geometry to construct the coarse meshes $\mathcal M_l=(V_{l}, E_{l}, \mathcal T_{l})$ and
prolongators  for level $l=1,2,\dots, L$. The vertices assigned to each coarse vertex form its smoothing block,
coupling the coarse representation and local solves through a single
partition.\begin{wrapfigure}{r}{0.5\columnwidth}
  \centering
  \includegraphics[width=\linewidth]{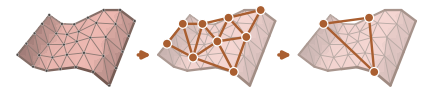}
\end{wrapfigure}

To build the coarse mesh $\mathcal M_{l+1}=(V_{l+1}, E_{l+1}, \mathcal T_{l+1})$ with the  partition of blocks in parallel, our pipeline consists of four stages for each level $0\leq l<L$. (1) At the \emph{coarse vertex selection} stage, we select the initial set of coarse vertices $V_{l+1}$. (2)  For the \emph{block partition} stage the fine vertices in $V_l$ are partitioned into $N_{l+1}$ disjoint subsets $\{B_v\}_{v\in V_{l+1}}$ where $B_v$ is the subset containing $v\in V_{l+1}$. The $V_{l+1}$ may be refined several rounds due to oversized blocks in this stage. Finally, the output is a mapping $f_l:V_l\to V_{l+1}$ such that $v\in B_{f_l(v)}$, is stored by indices. (3) An \emph{edge projection} stage is performed to create $E_{l+1}$. The coarser level mesh $\mathcal M_{l+1}$ is finalized in this stage. (4) Finally for the \emph{prolongator initialization} stage, we compute the prolongator $\bP_l$.

\begin{algorithm}[t]
\caption{Parallel Hierarchy and Block Construction}
\label{alg:coarsening-space-construction}
\begin{algorithmic}[1]
\renewcommand{\algorithmicrequire}{\textbf{Input:}}
\renewcommand{\algorithmicensure}{\textbf{Output:}}
\REQUIRE Fine mesh $\mathcal M_l$, vertex positions
$\{\mathbf p_{l,u}\}_{u\in V_l}$, initial level matrix
$\bA_l^{(0)}$ and its structural pattern,
block capacity $m_{\rm agg}$
\ENSURE Coarse mesh $\mathcal M_{l+1}$, block assignment $f_l$,
prolongator $\bP_l$, restriction $\bR_l$, and structural
coarse matrix pattern
\STATE Select initial coarse vertices $V_{l+1}$
using parallel MIS
\REPEAT
    \STATE Compute $f_l$ and mesh-edge distances by parallel
    multi-source propagation from $V_{l+1}$
    \STATE Form $B_v\gets\{u\in V_l:f_l(u)=v\}$
    for each $v\in V_{l+1}$
    \FOR{each block $B_v$ with $|B_v|>m_{\rm agg}$ in parallel}
        \STATE Select $\lceil |B_v|/m_{\rm agg}\rceil-1$
        additional coarse vertices from the farthest
        vertices in $B_v\setminus V_{l+1}$
    \ENDFOR
    \STATE Add the selected vertices to $V_{l+1}$
\UNTIL{every block satisfies the capacity bound}
\STATE Project crossing fine edges to construct $E_{l+1}$
\STATE Form $\mathcal M_{l+1}$ 
\STATE Construct $\bP_l$ using normalized
inverse-distance interpolation
\STATE Store $\bR_l=\bP_l^\top$ and compute the structural
pattern of $\bP_l^\top\bA_l^{(0)}\bP_l$
\RETURN $\mathcal M_{l+1}$, $f_l$, $\bP_l$, $\bR_l$,
and the coarse matrix pattern
\end{algorithmic}
\end{algorithm}
\begin{figure*}[t]
  \centering
  \includegraphics[width=0.95\textwidth]{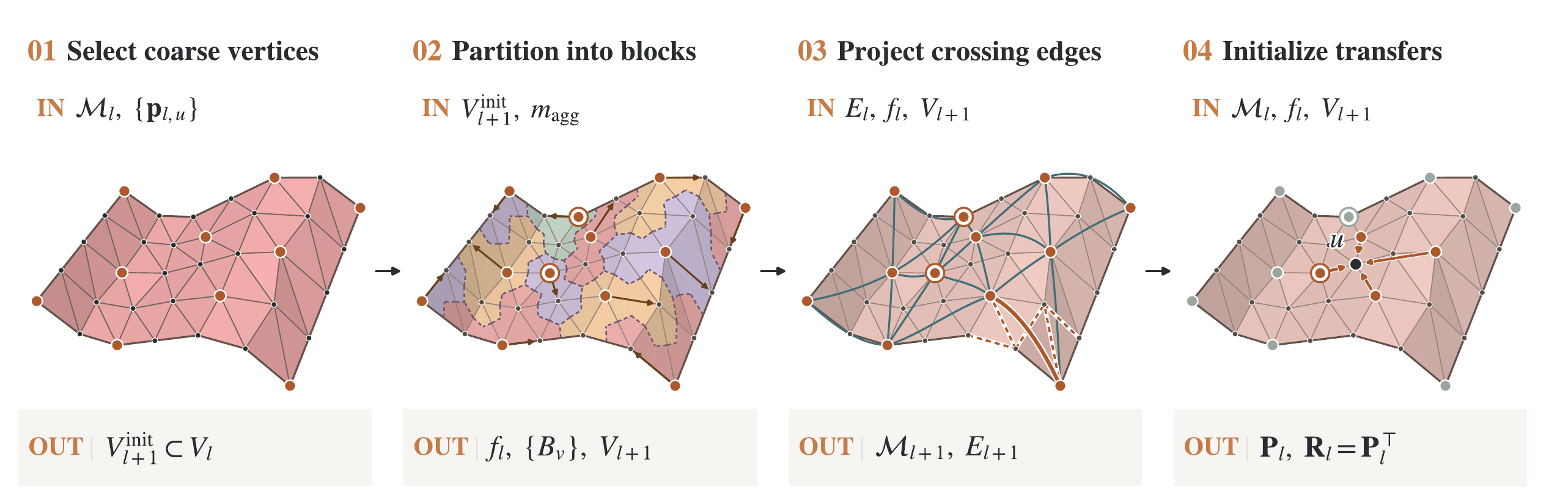}
  \caption{Parallel construction of one geometric hierarchy level.
  (1) Select coarse vertices to define the next mesh level.
  (2) Assign fine vertices to coarse vertices by shortest-path
  distance along mesh edges, forming one smoothing block per
  coarse vertex.
  (3) Project edges crossing blocks to construct coarse mesh
  connectivity.
  (4) Initialize the prolongator using normalized inverse-distance
  interpolation over local coarse vertex supports.}
  \Description{Four stages of parallel hierarchy construction:
  coarse vertex selection, block partition, edge projection,
  and prolongator initialization. Each coarse vertex represents
  one smoothing block of fine vertices.}
  \label{fig:method_hierarchy}
\end{figure*}

\paragraph{1. Coarse vertex selection.}
A parallel maximal independent set (PMIS) algorithm is used to select the coarse vertex set $V_{l+1}$ from $V_l$.
Each active vertex receives a deterministic priority based on
hashing its index and hierarchy level.
Each round selects local priority maxima as coarse vertices
and removes their conflict neighborhoods from the active set.
This is independent of matrix coefficients.
We also support coefficient-aware AHAT-D2
coarsening~\cite{naumov2015amgx}.
It identifies strong connections from the initial operator
coefficients at each level and applies PMIS using priorities
based on the number of incoming strong connections and a
deterministic vertex hash.
Both constructions support the capacity refinement described below,
which adds coarse vertices before interpolation is constructed.

\paragraph{2. Block partition.}
In this stage, we assign each fine vertex to its nearest coarse
vertex by shortest-path distance along mesh edges, weighted by
their Euclidean lengths. 
This distance is therefore the geodesic distance on the mesh wireframe,
providing a computationally efficient approximation to geodesic
distance over the mesh.

The resulting mapping $f_l:V_l\to V_{l+1}$ defines the block partition
\begin{equation}
  B_v=\{u\in V_l:f_l(u)=v\},
  \qquad v\in V_{l+1}.
  \label{eq:method_shared_partition}
\end{equation}

To find the shortest paths in parallel, each vertex $u$ maintains a label $(d_{l,u},f_l(u))$, where
$d_{l,u}$ is its current path distance. All coarse vertices propagate concurrently.
In each round, vertices whose labels improved in the preceding
round propose updates across their incident edges.
For an edge $(u,w)\in E_l$, the candidate distance is
\begin{equation}
  \widehat d
  =d_{l,u}+\|\mathbf p_{l,u}-\mathbf p_{l,w}\|_2.
  \label{eq:method_distance_candidate}
\end{equation}
If $\widehat d<d_{l,w}$, an atomic replacement updates the
label at $w$ to $(\widehat d,f_l(u))$ and schedules $w$ for
the next round. Propagation terminates when no label improves.

To bound the block solve size, each oversized block $B_v$
requests $\lceil |B_v|/m_{\rm agg}\rceil-1$ additional coarse
vertices.
We select vertices in $B_v\setminus V_{l+1}$ with the largest
current path distances and repeat the assignment using the
augmented coarse vertex set until $|B_v|\le m_{\rm agg}$
for every block.
The final partition contains one smoothing block per coarse
vertex.

Every accepted label represents a path from a coarse vertex.
At termination, no edge can shorten a label, so the assignment
recovers multi-source shortest paths in exact arithmetic.

This construction is inspired by the graph Voronoi-based coarsening
of prior work~\cite{wiersma2023gravo,padilla2026gravo}, whose CPU
pipelines construct a Voronoi graph sequentially and do not use it
for block-Jacobi smoothing. Our parallel construction produces a block partition that serves both the edge projection for constructing $\mathcal M_{l+1}$ and the block-Jacobi solves.

\paragraph{3. Edge projection.} When the coarse vertex-aligned blocks are partitioned, the edges crossing blocks induce the coarse edges.
\begin{equation}
  \begin{aligned}
    E_{l+1}=\bigl\{&(f_l(u),f_l(w)): (u,w)\in E_l,\ f_l(u)\ne f_l(w)\bigr\}.
  \end{aligned}
  \label{eq:method_quotient}
\end{equation}
Sorting and deduplication produce the coarse connectivity. The distance weight on the constructed edge set is computed by the Euclidean distance of the vertex positions.
The resulting mesh supplies the geometric neighborhoods for
the next coarsening step.

\paragraph{4. Prolongator initialization.}
We construct $\bP_l\in\mathbb R^{N_l\times N_{l+1}}$,
with rows and columns indexed by fine and coarse vertices,
respectively.
Rows corresponding to coarse vertices inject exactly,
with $(\bP_l)_{vv}=1$ and all other entries zero.
For each remaining vertex $u\in V_l\setminus V_{l+1}$,
the candidate coarse vertices are
\begin{equation}
  \mathcal C_{l,u}^{\rm cand}
  =
  \{f_l(u)\}
  \cup
  \{f_l(w):(u,w)\in E_l\}.
  \label{eq:supp_candidate_set}
\end{equation}
We retain up to four nearest candidates in Euclidean distance
to form the interpolation support
$\mathcal C_{l,u}\subseteq V_{l+1}$.
For coarse vertices, $\mathcal C_{l,v}=\{v\}$.
The complete support is
\begin{equation}
  \operatorname{supp}(\bP_l)
  =
  \{(u,v)\in V_l\times V_{l+1}:v\in\mathcal C_{l,u}\}.
  \label{eq:method_interpolation_support}
\end{equation}

For $u\in V_l\setminus V_{l+1}$ and
$v\in\mathcal C_{l,u}$, normalized inverse-distance
interpolation gives
\begin{equation}
  (\bP_l)_{uv}
  =
  \frac{\delta_{l,uv}^{-1}}
       {\sum_{w\in\mathcal C_{l,u}}\delta_{l,uw}^{-1}},
  \label{eq:method_idw}
\end{equation}
where $\delta_{l,uv}=\|\mathbf p_{l,u}-\mathbf p_{l,v}\|_2$.

The rows of $\bP_l$ sum to one, and exact injection ensures
full column rank.
Consequently,
$\bA_{l+1}^{(\nu)}=\bP_l^\top\bA_l^{(\nu)}\bP_l$
is SPD whenever $\bA_l^{(\nu)}$ is SPD.
We retain $\bP_l$ and $\bR_l=\bP_l^\top$ for repeated
V-cycle applications and use their fixed supports to determine
the structural coarse matrix pattern for numerical updates.

\textbf{Remark} This block-dependent interpolation strategy combines the sparsity of single-vertex interpolation \cite{naumov2015amgx, henson2002boomer} with the richer support of simplex-based interpolation \cite{wiersma2023gravo, padilla2026gravo},
balancing interpolation quality against numerical-update cost.
Since candidates depend on each vertex's neighborhood,
supports can differ within the same block. 
Vertices whose neighborhood lies entirely within one block have a single
nonzero, while vertices adjacent to other blocks can receive
contributions from multiple coarse vertices.
This allocation complements block smoothing, which reduces
intra-block errors, by concentrating richer interpolation near
block boundaries to represent cross-block variation.

\subsection{Block-Jacobi Smoothing}
\label{sec:decomposed-block-storage}

The hierarchy construction provides one smoothing block $B_v$
for each coarse vertex $v\in V_{l+1}$.
We use this partition for block-Jacobi smoothing, resolving
within-block couplings through independent local solves.
Larger blocks can improve convergence, but their benefit also
depends on the cost of matrix products and numerical updates.
We coordinate these operations through a decomposed matrix layout.
During V-cycle application, smoothing and residual updates require
only the block inverses and cross-block coefficients.

\paragraph{Block operator decomposition.}
For each level $0\le l<L$, the notation $\bA_l^{(\nu)}[B_v,B_v]$ selects the rows and
columns indexed by $B_v$.
We define
\begin{equation}
  \begin{aligned}
    \bD_l^{(\nu)}
      &=\operatorname{blockdiag}_{v\in V_{l+1}}
        \bA_l^{(\nu)}[B_v,B_v],\\
    \bE_l^{(\nu)}&=\bA_l^{(\nu)}-\bD_l^{(\nu)},
  \end{aligned}
  \label{eq:method_block_decomposition}
\end{equation}
where ``$\operatorname{blockdiag}$'' retains these principal blocks
and sets all \begin{wrapfigure}[6]{r}{0.33\columnwidth}
  \centering
  \includegraphics[width=\linewidth]{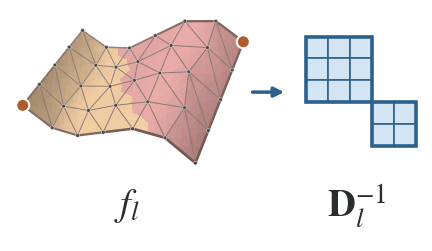}
\end{wrapfigure}cross-block entries to zero.
Both matrices belong to $\mathbb R^{N_l\times N_l}$.
Each principal block is SPD and has size
$|B_v|\times|B_v|$.
The inverse $(\bD_l^{(\nu)})^{-1}$ therefore applies independent
local solves, while $\bE_l^{(\nu)}$ contains the cross-block
coefficients.

\paragraph{Iterate and residual updates.}
Using the level variables from \autoref{eq:bg_v_cycle}, let
$\mathbf r_l=\mathbf b_l-\bA_l^{(\nu)}\mathbf x_l$ denote
the internal residual.
At the finest level, $\mathbf b_0=\mathbf r_k^{(\nu)}$ is
the input PCG residual.
A damped block-Jacobi smoother updates the correction as
\begin{equation}
  \mathbf x_l
  \leftarrow \mathbf x_l+\omega_l^{(\nu)}
    (\bD_l^{(\nu)})^{-1}\mathbf r_l.
  \label{eq:weighted-block-jacobi-correction}
\end{equation}
Substituting $\bA_l^{(\nu)}=\bD_l^{(\nu)}+\bE_l^{(\nu)}$
gives the equivalent iterate form
\begin{equation}
  \mathbf x_l^+
  =(1-\omega_l^{(\nu)})\mathbf x_l
   +\omega_l^{(\nu)}(\bD_l^{(\nu)})^{-1}
     (\mathbf b_l-\bE_l^{(\nu)}\mathbf x_l).
  \label{eq:method_block_iterate}
\end{equation}
The identity $\bA_l^{(\nu)}(\bD_l^{(\nu)})^{-1}
  =\mathbf I+\bE_l^{(\nu)}(\bD_l^{(\nu)})^{-1}$
also yields 
\begin{equation}
  \mathbf r_l\leftarrow(1-\omega_l^{(\nu)})\mathbf r_l
   -\omega_l^{(\nu)}\bE_l^{(\nu)}
     (\bD_l^{(\nu)})^{-1}\mathbf r_l.
  \label{eq:supp_residual_derivation}
\end{equation}

Each pre-smoothing sequence starts from $\mathbf x_l=\mathbf0$
and $\mathbf r_l=\mathbf b_l$.
Introducing the local correction $\mathbf y_l$ allows each
smoothing iteration to advance the iterate and residual together.
The first smoothing iteration therefore gives
$\mathbf x_l=\omega_l^{(\nu)}(\bD_l^{(\nu)})^{-1}\mathbf b_l$.
Subsequent smoothing iterations use the same recurrence, and the final residual
is restricted as $\mathbf b_{l+1}=\bR_l\mathbf r_l$.
After prolongating the coarse correction, post-smoothing uses
\autoref{eq:method_block_iterate}.
Both forms require only $(\bD_l^{(\nu)})^{-1}$ and
$\bE_l^{(\nu)}$, eliminating full level-matrix products
from smoothing and residual propagation.

\paragraph{Coefficient and inverse storage.}
The updates of matrix coefficients require the original coefficients in
$\bD_l^{(\nu)}$, in addition to $\bE_l^{(\nu)}$.
At each smoothing level, we store each block $B_v$ in a
  dense array of $|B_v|^2$ values. During numerical updates,
  these arrays are initialized with the current within-block
  coefficients, including zeros at structurally absent entries.
  The coefficients are first used for the coarse matrix assembly.
  Once this assembly is complete, we invert each block
  in place using the parallel Gauss-Jordan strategy employed
  in GPU MAS implementations~\cite{wu2022mas}, overwriting
  its coefficients with its inverse. Thus, coefficient
  assembly and inverse application reuse the same block
  storage, without a separate persistent inverse array.
  The next numerical update refills these arrays with
  the new coefficients before repeating assembly and inversion.  Cross-block coefficients in $\bE_l^{(\nu)}$ use compressed
  sparse row (CSR) storage containing both entries of each
  symmetric entry for fast operation at the \emph{apply} stage. This separate $\bD_l^{(\nu)}$ and $\bE_l^{(\nu)}$ layout avoids duplicating full CSR storage on intermediate levels while preserving the coefficients required by later updates.

\paragraph{Prepared Coarse Matrix assembly.}
Fixed prolongators and structural matrix patterns allow the coarse matrix assembly to be prepared once.
For fine vertices $i,j\in V_l$ with $i\ge j$ and coarse vertices
$u,v\in V_{l+1}$, define the scalar weight
$\gamma_{l,ij}^{uv}\in\mathbb R$ by
\begin{equation}
  \gamma_{l,ij}^{uv}=
  \begin{cases}
    (\bP_l)_{iu}(\bP_l)_{jv}
      +(\bP_l)_{ju}(\bP_l)_{iv},&i>j,\\
    (\bP_l)_{iu}(\bP_l)_{iv},&i=j.
  \end{cases}
  \label{eq:method_galerkin_weights}
\end{equation}
Vertex comparisons use their stored indices.
The off-diagonal case combines both orientations of a symmetric
fine entry.
Let $\operatorname{pat}(\bA_l)$ denote the fixed set of
structurally stored matrix indices.
For each coarse entry $(u,v)$, define its contribution set
$\mathcal L_{l,uv}\subseteq V_l\times V_l$ as
\begin{equation}
  \begin{aligned}
    \mathcal L_{l,uv}=\{(i,j)\in\operatorname{pat}(\bA_l):\quad\\
      i\ge j,\ \gamma_{l,ij}^{uv}\ne0\}.
  \end{aligned}
  \label{eq:method_galerkin_support}
\end{equation}
The numerical coarse coefficient is
\begin{equation}
  (\bA_{l+1}^{(\nu)})_{uv}
  =\sum_{(i,j)\in\mathcal L_{l,uv}}
     \gamma_{l,ij}^{uv}(\bA_l^{(\nu)})_{ij}.
  \label{eq:method_prepared_galerkin}
\end{equation}
At the \emph{setup} stage, we store each contribution set as a list of source addresses
in the coefficient blocks or cross-block CSR arrays, together
with its fixed weights and destination address.
During an update, one reduction per destination gathers the
current coefficients and writes their weighted sum directly
into $\bD_{l+1}^{(\nu)}$ or $\bE_{l+1}^{(\nu)}$.
Cross-block results are also written to their symmetric CSR
positions.
At $l+1=L$, results are written to the coarsest CSR matrix.
The same procedure handles within-block and cross-block outputs
regardless of where their source coefficients are stored.
It avoids repeated sparsity searches and floating-point atomic
accumulation.

Since we perform in-place block inversions, we should manage the execution order carefully in the \emph{update} stage.
Starting from $\bA_0^{(\nu)}$, the prepared reductions refresh
the coefficients throughout the hierarchy.
The block inverses and coarsest solver are then refreshed.

Our progressive ablation shows that larger smoothing blocks
reduce iteration counts, while the decomposed storage and
prepared updates are needed to translate this improvement into
lower total solve time (\autoref{sec:analysis-components}).


\textbf{Remark} The decomposed layout is particularly suited to V-cycles
  with paired block-Jacobi pre- and post-smoothing.
   Relative to the
  local block-inverse application used in MAS, the additional
  operator required for smoothing and residual propagation
  is only $\bE_l^{(\nu)}$, together with its matrix-vector
  products. These products account for inter-block coupling
  and, together with coarse level corrections, substantially
  improve convergence over MAS preconditioning.
\subsection{Damping Adapted to Evolving Linear Systems}
\label{sec:method_spectrum}

The block-Jacobi updates in
\autoref{eq:weighted-block-jacobi-correction}
use a damping weight $\omega_l^{(\nu)}$ at each smoothing level.
As matrix coefficients evolve, we adapt these weights using
an estimate of the largest eigenvalue of
\begin{equation}
  \bH_l^{(\nu)}
  =(\bD_l^{(\nu)})^{-1}\bA_l^{(\nu)}
  =\mathbf I+(\bD_l^{(\nu)})^{-1}\bE_l^{(\nu)}.
  \label{eq:method_power_operator}
\end{equation}
Applying this operator reuses the block inverses and cross-block
coefficients prepared for smoothing.
 
During \emph{setup}, we perform many power iterations on
$\bH_l^{(0)}$ to initialize a unit vector
$\mathbf v_l\in\mathbb R^{N_l}$ approximating its dominant
eigenvector. We define the corresponding estimated eigenvalue as
$
  \widehat\lambda_l^{(\nu)}
  =\|\bH_l^{(\nu)}\mathbf v_l\|_2$.
As coefficients evolve, we retain this vector and advance it
by one power iteration during each \emph{update} stage,
$
  \mathbf v_l\leftarrow
  \frac{\bH_l^{(\nu)}\mathbf v_l}
       {\widehat\lambda_l^{(\nu)}}.$
This reuse tracks gradual spectral changes across consecutive
linear solves at the cost of one operator application per level. We empirically choose $\omega_{l}^{(\nu)} = \frac{\vartheta}{\widehat\lambda_l^{(\nu)}}$ where $0<\vartheta<1$ is the spectrum margin.
The updated damping weights remain fixed throughout each
PCG solve.

\textbf{Remark} The light-weight Jacobi coefficient adaptation strategy introduced in this section is specifically tailored to sequences of evolving linear systems. GravoTet~\cite{padilla2026gravo} similarly uses power
  iteration for spectral estimation, with a fixed budget
  of iterations. Conversely, we retain the estimated dominant eigenvector and refine it as the coefficients change. Fewer power
  iterations per \emph{update} are typically sufficient
  to track gradual spectral changes, reducing the cost of
  adapting the damping weights across the sequence.
\subsection{Coarsest Solve and Re-ordering}
\label{sec:coarse_solve_ordering}
We prepare a GPU direct solver for the coarsest system using block Cholesky elimination. The reordered coarsest   matrix admits the factorization $\bA_{L}^{(\nu)}=\bL\bL^{\top}$, where $\bL$ is lower triangular. Given the restricted right-hand side $\mathbf{b}_{L}$, the solver returns the  coarse correction $\bx_{L}=\mathbf{L}^{-\top}\mathbf{L}^{-1}\mathbf{b}_{L}$. The symbolic elimination structure depends only on the matrix sparsity pattern and is constructed once  during setup. When the coefficients change, numerical preparation refreshes the factorization and derives local inverse and coupling operators, allowing repeated coarse solves to use parallel matrix–vector products along the elimination tree. The mathematical derivation and implementation details are deferred to the the supplementary document.

\textbf{Remark}
  We develop this coarsest direct solver to avoid dependence on
  closed-source sparse direct solvers. Moreover, our smoother is also inspired by the block design of the sparse direct solve. Compared to the block-Jacobi smoother, the coarsest direct solver also organizes computation around blocks of unknowns, while accounting for inter-block coupling through direct
  elimination. This shared block-based
  organization also suggests a favor of direct
  solving on small meshes.

The vertices of the coarsest mesh $V_L$ should be therefore reordered according to the nested dissection elimination tree of the coarsest direct solver.
  The coarsest ordering also determines the data layout throughout the hierarchy. During \emph{setup}, we apply nested dissection to the coarsest sparsity  pattern, recursively separating the unknowns into elimination groups and ordering their interior unknowns before the separating unknowns. This ordering defines the dependencies of the coarsest elimination. We then propagate it from coarse to fine through the existing block mappings $f_l$ for $0\leq l<L$. Each coarse vertex determines the position of its associated fine-level smoothing block, whose vertices are stored contiguously. At every level, the same  permutation is applied to the rows and columns of the system matrix, while each transfer operator is reordered according to the orderings of its fine and coarse levels. These determine the global re-ordering of the original vertex set $V_0$. 
  
\textbf{Remark}  The relationship between ordering and block construction
  differs from ordering-based MAS implementations \cite{wu2022mas}, which
  first order the unknowns and then partition the blocks according to the ordering, producing \emph{ordering-induced blocks}.
  ParaBMG instead constructs \emph{partition-induced blocks} 
  from the mesh and derives the vertex ordering from these
  partitions.

\section{Benchmark and Evaluation}
We evaluate ParaBMG on sequences of evolving linear systems generated by our parallel implementations of standard geometry processing algorithms. We first introduce the benchmark problems and evaluation protocol, then compare ParaBMG with SOTA GPU iterative solvers. We further examine performance variation across the recorded systems, analyze the runtime breakdown, and compare with direct solvers.

\begin{figure*}[t]
    \centering
    \includegraphics[width=\textwidth]{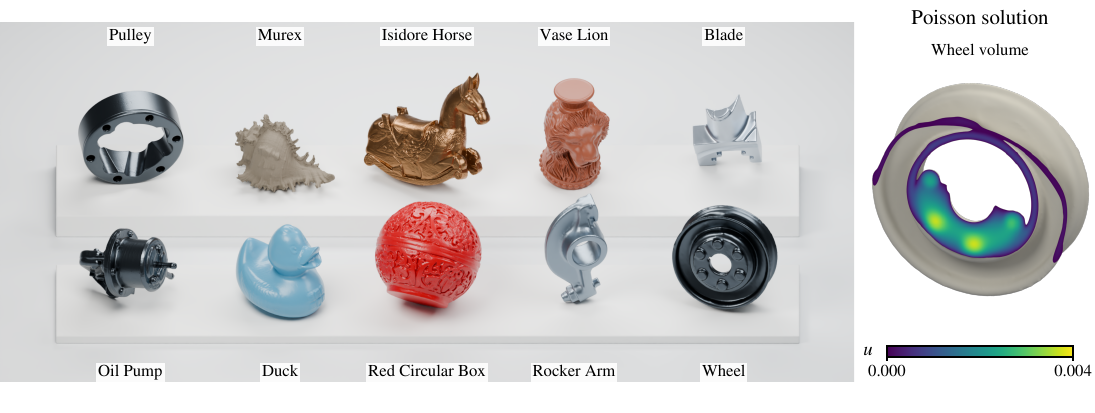}
    \caption{\textbf{Left:} the ten benchmark meshes sourced from previous works \cite{wiersma2023gravo}. The number of vertices for each mesh is given in \autoref{tab:evaluation-applications}. \textbf{Right:} one Poisson solution illustrated by the scalar field heatmap on a cutting plane.}
    \label{fig:mesh-gallery}
\end{figure*}
\begin{table*}[p]
\centering
\caption{Solver performance on ten meshes and four application problems. We report the average cost (ms) of one update and one solve defined in \autoref{eq:bg_sequence_cost}. Bold marks the lowest unrounded $T$ in each row. Speedup is $T_{\mathrm{baseline}}/T_{\mathrm{ours}}$. The background color of timing cells encodes runtime relative to the fastest baseline in each row: white marks the fastest baseline, red marks slower methods, and green marks faster methods, with stronger colors indicating larger differences.}
\label{tab:evaluation-applications}
\begin{tabular}{@{}llrrrrrrrrrr@{}}
\toprule
\multirow{2}{*}{Mesh} & \multirow{2}{*}{Problem} & \multirow{2}{*}{$N$} & \multicolumn{2}{c}{MAS} & \multicolumn{2}{c}{AMGX} & \multicolumn{2}{c}{BoomerAMG} & \multicolumn{2}{c}{GravoMG-GPU} & \multirow{2}{*}{Ours} \\
\cmidrule(lr){4-5}\cmidrule(lr){6-7}\cmidrule(lr){8-9}\cmidrule(lr){10-11}
 & & & $T$ (ms) & Speedup & $T$ (ms) & Speedup & $T$ (ms) & Speedup & $T$ (ms) & Speedup & \\
\midrule

\multirow{4}{*}{Wheel} & LDSE & 784,680 & \cellcolor[HTML]{FBE2E1}219.5 & $23.1\times$ & \cellcolor[HTML]{FFFFFF}52.9 & $5.6\times$ & \cellcolor[HTML]{FFFDFD}58.3 & $6.1\times$ & \cellcolor[HTML]{FDF1F1}104.3 & $11.0\times$ & \cellcolor[HTML]{DDF0E4}\textbf{9.5} \\

 & QHW & 784,680 & \cellcolor[HTML]{FADDDC}311.4 & $31.5\times$ & \cellcolor[HTML]{FFFFFF}58.8 & $5.9\times$ & \cellcolor[HTML]{FFFDFD}63.3 & $6.4\times$ & \cellcolor[HTML]{FDF3F3}104.6 & $10.6\times$ & \cellcolor[HTML]{DCF0E3}\textbf{9.9} \\

 & IMCF & 291,695 & \cellcolor[HTML]{FADAD9}135.4 & $29.4\times$ & \cellcolor[HTML]{FFFFFF}22.5 & $4.9\times$ & \cellcolor[HTML]{FEFAFA}28.1 & $6.1\times$ & \cellcolor[HTML]{F7C6C4}483.5 & $105.1\times$ & \cellcolor[HTML]{E0F1E6}\textbf{4.6} \\
 & DSTV & 291,695 & \cellcolor[HTML]{FEF6F6}32.6 & $10.4\times$ & \cellcolor[HTML]{FFFFFF}21.5 & $6.9\times$ & \cellcolor[HTML]{FFFEFE}22.6 & $7.2\times$ & \cellcolor[HTML]{F8CFCE}217.7 & $69.3\times$ & \cellcolor[HTML]{D9EEE1}\textbf{3.1} \\
\midrule

\multirow{4}{*}{Blade} & LDSE & 1,324,183 & \cellcolor[HTML]{FBE4E3}293.4 & $20.6\times$ & \cellcolor[HTML]{FFFFFF}78.9 & $5.5\times$ & \cellcolor[HTML]{FFFEFE}82.5 & $5.8\times$ & \cellcolor[HTML]{FDF0EF}164.9 & $11.6\times$ & \cellcolor[HTML]{DDF0E4}\textbf{14.3} \\

 & QHW & 1,324,183 & \cellcolor[HTML]{FADAD8}533.9 & $35.4\times$ & \cellcolor[HTML]{FFFDFD}95.5 & $6.3\times$ & \cellcolor[HTML]{FFFFFF}86.4 & $5.7\times$ & \cellcolor[HTML]{FDF2F1}165.1 & $11.0\times$ & \cellcolor[HTML]{DCF0E4}\textbf{15.1} \\

 & IMCF & 176,545 & \cellcolor[HTML]{FADEDD}85.1 & $24.9\times$ & \cellcolor[HTML]{FFFFFF}17.0 & $5.0\times$ & \cellcolor[HTML]{FEF8F8}23.6 & $6.9\times$ & \cellcolor[HTML]{FADBD9}99.3 & $29.1\times$ & \cellcolor[HTML]{DFF1E6}\textbf{3.4} \\
 & DSTV & 176,545 & \cellcolor[HTML]{FFFEFE}15.6 & $8.0\times$ & \cellcolor[HTML]{FFFFFF}15.0 & $7.7\times$ & \cellcolor[HTML]{FFFEFE}15.0 & $7.7\times$ & \cellcolor[HTML]{FCE9E8}43.5 & $22.4\times$ & \cellcolor[HTML]{D6EDDF}\textbf{1.9} \\
\midrule

\multirow{4}{*}{Vase Lion} & LDSE & 1,363,733 & \cellcolor[HTML]{FBE4E3}321.5 & $18.7\times$ & \cellcolor[HTML]{FFFFFF}87.4 & $5.1\times$ & \cellcolor[HTML]{FFFEFE}88.0 & $5.1\times$ & \cellcolor[HTML]{FDF0F0}181.0 & $10.5\times$ & \cellcolor[HTML]{DFF1E6}\textbf{17.2} \\

 & QHW & 1,363,733 & \cellcolor[HTML]{FADEDD}447.2 & $25.3\times$ & \cellcolor[HTML]{FFFEFE}91.6 & $5.2\times$ & \cellcolor[HTML]{FFFFFF}89.4 & $5.1\times$ & \cellcolor[HTML]{FDEEEE}202.6 & $11.5\times$ & \cellcolor[HTML]{DFF1E6}\textbf{17.6} \\

 & IMCF & 196,358 & \cellcolor[HTML]{FBE5E4}69.0 & $18.2\times$ & \cellcolor[HTML]{FFFFFF}19.4 & $5.1\times$ & \cellcolor[HTML]{FEF9F9}25.7 & $6.8\times$ & \cellcolor[HTML]{F9D6D5}140.1 & $37.0\times$ & \cellcolor[HTML]{DFF1E6}\textbf{3.8} \\
 & DSTV & 196,358 & \cellcolor[HTML]{FFFEFE}15.9 & $6.6\times$ & \cellcolor[HTML]{FFFFFF}14.9 & $6.2\times$ & \cellcolor[HTML]{FFFEFE}15.9 & $6.6\times$ & \cellcolor[HTML]{FCE7E6}47.7 & $19.7\times$ & \cellcolor[HTML]{DBEFE3}\textbf{2.4} \\
\midrule

\multirow{4}{*}{Duck} & LDSE & 1,393,355 & \cellcolor[HTML]{FBE1E0}371.7 & $21.0\times$ & \cellcolor[HTML]{FFFFFF}88.3 & $5.0\times$ & \cellcolor[HTML]{FFFEFE}91.6 & $5.2\times$ & \cellcolor[HTML]{FDEFEE}192.5 & $10.9\times$ & \cellcolor[HTML]{DFF1E6}\textbf{17.7} \\

 & QHW & 1,393,355 & \cellcolor[HTML]{FADEDD}479.2 & $26.7\times$ & \cellcolor[HTML]{FFFEFE}98.8 & $5.5\times$ & \cellcolor[HTML]{FFFFFF}95.1 & $5.3\times$ & \cellcolor[HTML]{FDEEED}219.8 & $12.3\times$ & \cellcolor[HTML]{DEF1E5}\textbf{17.9} \\

 & IMCF & 202,554 & \cellcolor[HTML]{FBE4E3}69.9 & $18.8\times$ & \cellcolor[HTML]{FFFFFF}19.0 & $5.1\times$ & \cellcolor[HTML]{FEFAFA}24.7 & $6.6\times$ & \cellcolor[HTML]{FADBDA}108.0 & $29.1\times$ & \cellcolor[HTML]{DFF1E6}\textbf{3.7} \\
 & DSTV & 202,554 & \cellcolor[HTML]{FFFEFE}13.4 & $6.3\times$ & \cellcolor[HTML]{FFFFFF}12.5 & $5.9\times$ & \cellcolor[HTML]{FEF7F7}18.1 & $8.5\times$ & \cellcolor[HTML]{FBE4E3}45.8 & $21.6\times$ & \cellcolor[HTML]{DCF0E3}\textbf{2.1} \\
\midrule

\multirow{4}{*}{Rocker Arm} & LDSE & 1,414,721 & \cellcolor[HTML]{FBE6E5}295.9 & $18.6\times$ & \cellcolor[HTML]{FFFFFF}86.1 & $5.4\times$ & \cellcolor[HTML]{FFFEFE}87.4 & $5.5\times$ & \cellcolor[HTML]{FDF0F0}176.6 & $11.1\times$ & \cellcolor[HTML]{DEF0E5}\textbf{15.9} \\

 & QHW & 1,414,721 & \cellcolor[HTML]{FADEDD}446.4 & $26.8\times$ & \cellcolor[HTML]{FFFCFC}103.0 & $6.2\times$ & \cellcolor[HTML]{FFFFFF}89.6 & $5.4\times$ & \cellcolor[HTML]{FDF0F0}184.9 & $11.1\times$ & \cellcolor[HTML]{DEF0E5}\textbf{16.7} \\

 & IMCF & 254,339 & \cellcolor[HTML]{FADDDC}104.1 & $25.5\times$ & \cellcolor[HTML]{FFFFFF}20.3 & $5.0\times$ & \cellcolor[HTML]{FEFAFA}25.9 & $6.3\times$ & \cellcolor[HTML]{FEFAFA}25.9 & $6.3\times$ & \cellcolor[HTML]{DFF1E6}\textbf{4.1} \\
 & DSTV & 254,339 & \cellcolor[HTML]{FEFAF9}19.7 & $8.2\times$ & \cellcolor[HTML]{FFFFFF}15.2 & $6.3\times$ & \cellcolor[HTML]{FFFDFD}16.6 & $6.9\times$ & \cellcolor[HTML]{FFFEFE}15.2 & $6.4\times$ & \cellcolor[HTML]{DAEFE2}\textbf{2.4} \\
\midrule

\multirow{4}{*}{Murex} & LDSE & 5,299,495 & \cellcolor[HTML]{FBE4E3}1155.6 & $22.8\times$ & \cellcolor[HTML]{FFFCFC}358.8 & $7.1\times$ & \cellcolor[HTML]{FFFFFF}316.6 & $6.3\times$ & \cellcolor[HTML]{FCEAEA}869.5 & $17.2\times$ & \cellcolor[HTML]{DBEFE3}\textbf{50.6} \\

 & QHW & 5,299,495 & \cellcolor[HTML]{FAD8D7}2490.2 & $41.6\times$ & \cellcolor[HTML]{FEF7F7}563.3 & $9.4\times$ & \cellcolor[HTML]{FFFFFF}379.0 & $6.3\times$ & \cellcolor[HTML]{FCECEC}932.7 & $15.6\times$ & \cellcolor[HTML]{DAEFE2}\textbf{59.8} \\

 & IMCF & 1,647,843 & \cellcolor[HTML]{F7C9C7}833.6 & $61.8\times$ & \cellcolor[HTML]{FFFFFF}59.3 & $4.4\times$ & \cellcolor[HTML]{FFFDFD}65.2 & $4.8\times$ & \cellcolor[HTML]{FCEDEC}142.0 & $10.5\times$ & \cellcolor[HTML]{E2F2E8}\textbf{13.5} \\
 & DSTV & 1,647,843 & \cellcolor[HTML]{FDF2F2}64.3 & $9.6\times$ & \cellcolor[HTML]{FFFFFF}34.7 & $5.2\times$ & \cellcolor[HTML]{FEF9F9}45.5 & $6.8\times$ & \cellcolor[HTML]{FDF3F3}61.2 & $9.2\times$ & \cellcolor[HTML]{DEF1E5}\textbf{6.7} \\
\midrule

\multirow{4}{*}{\shortstack[l]{Isidore\\Horse}} & LDSE & 5,770,814 & \cellcolor[HTML]{F9D6D5}2657.5 & $45.9\times$ & \cellcolor[HTML]{FEFBFB}439.4 & $7.6\times$ & \cellcolor[HTML]{FFFFFF}367.8 & $6.4\times$ & \cellcolor[HTML]{FCE9E8}1069.2 & $18.5\times$ & \cellcolor[HTML]{DAEFE2}\textbf{57.8} \\

 & QHW & 5,770,814 & \cellcolor[HTML]{F8CBC9}5963.2 & $91.0\times$ & \cellcolor[HTML]{FEF7F7}700.4 & $10.7\times$ & \cellcolor[HTML]{FFFFFF}471.1 & $7.2\times$ & \cellcolor[HTML]{FBE3E2}1861.5 & $28.4\times$ & \cellcolor[HTML]{D8EEE0}\textbf{65.5} \\

 & IMCF & 1,116,358 & \cellcolor[HTML]{F8D0CF}431.2 & $42.5\times$ & \cellcolor[HTML]{FFFFFF}44.3 & $4.4\times$ & \cellcolor[HTML]{FFFCFC}50.2 & $5.0\times$ & \cellcolor[HTML]{FADBD9}260.1 & $25.7\times$ & \cellcolor[HTML]{E2F2E8}\textbf{10.1} \\
 & DSTV & 1,116,358 & \cellcolor[HTML]{FDEEED}53.6 & $10.2\times$ & \cellcolor[HTML]{FFFFFF}23.4 & $4.4\times$ & \cellcolor[HTML]{FEF4F4}39.3 & $7.4\times$ & \cellcolor[HTML]{FBE0DF}107.7 & $20.4\times$ & \cellcolor[HTML]{E1F2E8}\textbf{5.3} \\
\midrule

\multirow{4}{*}{Pulley} & LDSE & 6,522,585 & \cellcolor[HTML]{F9D7D6}2891.1 & $48.0\times$ & \cellcolor[HTML]{FEFBFA}511.5 & $8.5\times$ & \cellcolor[HTML]{FFFFFF}412.7 & $6.9\times$ & \cellcolor[HTML]{FCEBEA}1082.1 & $18.0\times$ & \cellcolor[HTML]{D9EEE1}\textbf{60.2} \\

 & QHW & 6,522,585 & \cellcolor[HTML]{F7CAC8}6740.3 & $96.3\times$ & \cellcolor[HTML]{FEF9F8}685.2 & $9.8\times$ & \cellcolor[HTML]{FFFFFF}499.7 & $7.1\times$ & \cellcolor[HTML]{FDEEEE}1128.3 & $16.1\times$ & \cellcolor[HTML]{D8EEE1}\textbf{70.0} \\

 & IMCF & 582,324 & \cellcolor[HTML]{FADAD8}183.9 & $34.3\times$ & \cellcolor[HTML]{FFFFFF}29.8 & $5.6\times$ & \cellcolor[HTML]{FFFDFD}33.4 & $6.2\times$ & \cellcolor[HTML]{FBE2E0}125.1 & $23.3\times$ & \cellcolor[HTML]{DDF0E4}\textbf{5.4} \\
 & DSTV & 582,324 & \cellcolor[HTML]{FEF8F7}27.6 & $9.0\times$ & \cellcolor[HTML]{FFFFFF}19.2 & $6.3\times$ & \cellcolor[HTML]{FFFCFB}22.7 & $7.4\times$ & \cellcolor[HTML]{FBE6E5}66.1 & $21.6\times$ & \cellcolor[HTML]{DBEFE3}\textbf{3.1} \\
\midrule

\multirow{4}{*}{Oil Pump} & LDSE & 6,553,054 & \cellcolor[HTML]{F9D8D6}2799.6 & $42.9\times$ & \cellcolor[HTML]{FEFAFA}524.1 & $8.0\times$ & \cellcolor[HTML]{FFFFFF}411.3 & $6.3\times$ & \cellcolor[HTML]{FCE8E7}1261.3 & $19.3\times$ & \cellcolor[HTML]{DAEFE2}\textbf{65.2} \\

 & QHW & 6,553,054 & \cellcolor[HTML]{F8CFCD}5270.6 & $72.1\times$ & \cellcolor[HTML]{FEFAFA}646.7 & $8.8\times$ & \cellcolor[HTML]{FFFFFF}509.8 & $7.0\times$ & \cellcolor[HTML]{FCEDEC}1221.3 & $16.7\times$ & \cellcolor[HTML]{D8EEE1}\textbf{73.1} \\

 & IMCF & 512,051 & \cellcolor[HTML]{F9D4D2}254.0 & $31.3\times$ & \cellcolor[HTML]{FFFFFF}30.6 & $3.7\times$ & \cellcolor[HTML]{FEFBFB}36.8 & $4.5\times$ & \cellcolor[HTML]{F7C7C5}468.0 & $57.6\times$ & \cellcolor[HTML]{E5F4EA}\textbf{8.1} \\
 & DSTV & 512,051 & \cellcolor[HTML]{FEF5F5}29.5 & $10.4\times$ & \cellcolor[HTML]{FFFFFF}18.3 & $6.4\times$ & \cellcolor[HTML]{FFFCFB}21.6 & $7.6\times$ & \cellcolor[HTML]{FBE0DF}80.9 & $28.5\times$ & \cellcolor[HTML]{DAEFE2}\textbf{2.8} \\
\midrule

\multirow{4}{*}{\shortstack[l]{Red Circular\\Box}} & LDSE & 6,600,234 & \cellcolor[HTML]{F9D7D5}3247.9 & $47.8\times$ & \cellcolor[HTML]{FEFBFB}562.5 & $8.3\times$ & \cellcolor[HTML]{FFFFFF}459.8 & $6.8\times$ & \cellcolor[HTML]{FCEBEA}1228.9 & $18.1\times$ & \cellcolor[HTML]{D9EEE1}\textbf{67.9} \\

 & QHW & 6,600,234 & \cellcolor[HTML]{F9D2D0}4934.8 & $66.7\times$ & \cellcolor[HTML]{FEFBFB}669.0 & $9.0\times$ & \cellcolor[HTML]{FFFFFF}541.6 & $7.3\times$ & \cellcolor[HTML]{FCEBEA}1450.4 & $19.6\times$ & \cellcolor[HTML]{D8EEE0}\textbf{74.0} \\

 & IMCF & 692,972 & \cellcolor[HTML]{FADEDD}169.0 & $28.1\times$ & \cellcolor[HTML]{FFFFFF}33.7 & $5.6\times$ & \cellcolor[HTML]{FFFCFC}38.2 & $6.4\times$ & \cellcolor[HTML]{FADDDB}179.1 & $29.8\times$ & \cellcolor[HTML]{DDF0E4}\textbf{6.0} \\
 & DSTV & 692,972 & \cellcolor[HTML]{FEF6F6}37.3 & $7.8\times$ & \cellcolor[HTML]{FFFFFF}24.6 & $5.1\times$ & \cellcolor[HTML]{FFFCFC}28.3 & $5.9\times$ & \cellcolor[HTML]{F8CFCD}258.9 & $53.9\times$ & \cellcolor[HTML]{DFF1E6}\textbf{4.8} \\
\bottomrule
\end{tabular}
\end{table*}

\subsection{Experimental Protocol}

Our benchmark dataset comprises four geometry processing pipelines. Local deformation for interactive shape editing
(LDSE)~\cite{chen2023ldef} deforms a volumetric mesh under prescribed constraints. The variational quasi-harmonic weight (QHW)~\cite{wang2021qhw} computes volumetric weights for shape deformation. Inverse optimization through implicit mean curvature flow (IMCF)~\cite{dodik2026iskra} seeks a surface whose flow step matches a target shape. The directional shape total variation (DSTV)~\cite{brokman2024dstv} filters a surface along prescribed directions. LDSE and QHW use tetrahedral volume meshes, whereas IMCF and DSTV use triangular surface meshes. Our CUDA implementations generate 40 sequences of linear systems, involving 40,987 instances of linear systems in total. All experiments use relative residual error $10^{-6}$ as the stopping criterion of the iterative solving methods.

\paragraph{Mesh Preparation} We select meshes from the collection used by
\citet{wiersma2023gravo}. After excluding models depicting human faces or featuring religious or cultural associations, we retain the ten largest remaining meshes. We apply modest refinement before tetrahedralization and retain the surface triangle meshes for IMCF and DSTV. The surface meshes contain 176,545-1,647,843 vertices,
and the tetrahedral meshes contain 784,680-6,600,234 vertices. \autoref{fig:mesh-gallery} shows the meshes, and
\autoref{tab:evaluation-applications} gives their vertex counts.

\paragraph{Hardware} 
The  solver benchmarks in \autoref{tab:evaluation-applications} run on one NVIDIA A100 PCIe GPU and CUDA 12.9. The interactive demonstrations in \autoref{sec:applications} use a separate workstation with an NVIDIA GeForce RTX 5080 GPU and CUDA 13.2.

\paragraph{Evaluation Metric}
We follow the average timing cost \autoref{eq:bg_sequence_cost} as the evaluation metric, which averages the cost of the \emph{update} and \emph{apply} stage.

\paragraph{Compared Solvers.}
We compare ParaBMG with multilevel additive Schwarz (MAS), AMGX,  BoomerAMG, and GravoMG-GPU. These baselines cover SOTA linear solvers in elasticity simulation, SOTA general-purpose algebraic multigrid, and GPU accelerated geometric multigrid approaches.
MAS follows \citet{wu2022mas}, adapted to scalar
$N\times N$ systems. It uses single-precision (FP32) local Cholesky factorizations, and double-precision (FP64) outer PCG.
For the geometric multigrid comparison, we implement both the
surface mesh version of \citet{wiersma2023gravo} and the volumetric mesh version of \citet{padilla2026gravo} in CUDA. Together, these
implementations are denoted GravoMG-GPU. The surface implementation retains Gauss-Seidel smoothing through CUDA sparse triangular solves. The volumetric implementation retains Chebyshev smoothing with eigenvalue estimation. Both adopt the V-cycle multigrid as a fixed-point iterative solving method rather than using it as a preconditioner. This preserves the original behavior of the CPU implementations  while enabling comparison on the same GPU. For AMGX and BoomerAMG, we compare with their open-sourced GPU implementations.

\paragraph{Baseline Configuration.}
AMGX and BoomerAMG offer several choices of coarsening,
interpolation, and smoothing parameters. To account for their
effect on performance, we test multiple parameter configurations for each sequence of evolving linear systems and report the best measured cost among the tested valid configurations. A configuration specifies one combination of solver parameters, and the selection is based on timing for each sequence of evolving linear system. The selected configuration remains fixed for all linear systems in the sequence.  For AMGX, we test classical AMG with PMIS coarsening and D2 interpolation, and aggregation AMG with SIZE\_2 aggregation. Both use $l_1$-Jacobi smoothing. The aggregation configurations vary the relaxation factor between 0.6 and 0.8 and use one or two smoothing sweeps. For BoomerAMG, we retain PMIS coarsening and $l_1$-Jacobi smoothing while varying the strength threshold and interpolation width.

\subsection{Overall Performance}

\begin{figure}[!htb]
  \centering
  \includegraphics[width=0.8\columnwidth]{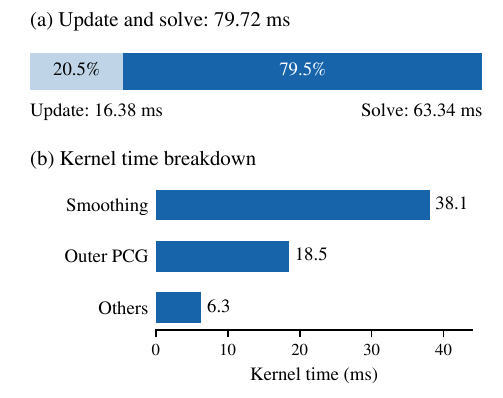}
  \caption{\textbf{Runtime breakdown on one typical Pulley/LDSE system.}
  Top: one \emph{update} and one PCG \emph{solve},
  totaling 79.72\,ms and excluding one-time setup.
  Bottom: kernel-time allocation from a separate profile of the same system.}
  \label{fig:runtime-breakdown}
  \Description{Two bar charts show update and solve time, followed
  by the profiled solve's kernel-time allocation. Block smoothing,
  outer PCG operations, and multigrid residual formation dominate
  the kernel time.}
\end{figure}

\label{sec:primary_benchmark}

\begin{figure}[!htb]
  \centering
  \includegraphics[width=\columnwidth]{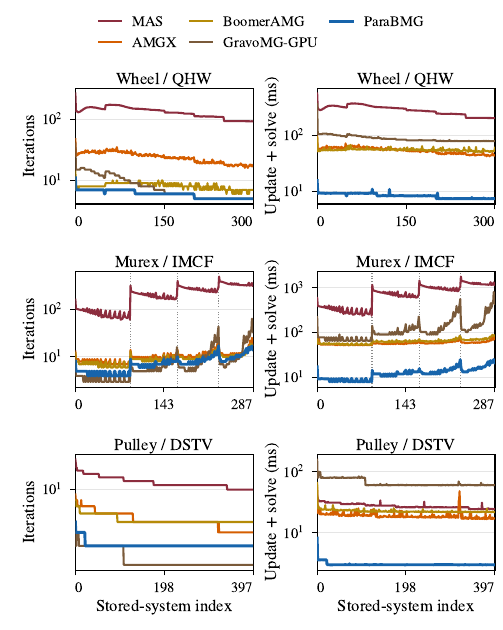}
  \caption{ Complete Wheel/QHW, Murex/IMCF, and Pulley/DSTV histories. Left: iteration count of the compared iterative solvers. Right: the time cost of the solvers.}
  \label{fig:analysis-stored-trajectories}
  \Description{.}
\end{figure}

\autoref{tab:evaluation-applications} compares average solving costs across 40 sequences of evolving linear systems.
ParaBMG has a lower cost $T$ than AMGX, BoomerAMG,
and GravoMG-GPU on every sequence. Relative to the fastest of
these three baselines on each sequence, the reported costs give speedups of $3.8\times$ to $7.7\times$.

\paragraph{Comparison Details.} ParaBMG combines low iteration counts with efficient execution. Across all 20 volumetric sequences generated by LDSE and QHW, it requires fewer iterations than any of the baselines, averaging 6.3-8.7 iterations per linear system. On triangular surface meshes, GravoMG-GPU requires fewer iterations for 9 out of 20 sequences, but ParaBMG remains $6\times$–$105\times$ faster overall. General-purpose AMG solvers, including AMGX and BoomerAMG, consistently deliver relatively high performance among the baselines. Compared with ParaBMG, however, they incur higher hierarchy update costs, require more PCG iterations, and have higher per-iteration costs. Our customized MAS implementation consistently requires approximately an order of magnitude more PCG iterations than ParaBMG, despite its lower per-iteration cost. This suggests that multigrid may be better suited to the Poisson-type systems in our benchmarks than MAS.

\paragraph{Variation across Sequences.}
We next examine how ParaBMG's performance varies within the
recorded sequences. For each sequence, we compute the iteration
coefficient of variation (CV), defined as the sample standard
deviation divided by the mean, and the ratio of the 95th percentile
of the combined update and apply time to its median.

QHW and DSTV exhibit relatively modest variation, with median
iteration CVs of 13.1\% and 11.2\% and corresponding time ratios
of 1.10 and 1.15. LDSE and IMCF show greater variation, with
iteration CVs of 31.4\% and 28.5\% and time ratios of 1.57 and
1.30, respectively. For LDSE, each sequence includes ten
independent editing cases, so these statistics capture both
variation between cases and the transition from an initial
solve to subsequent solves that reuse the previous solution.
IMCF also includes independently initialized timestep groups and
line search trial systems. These results show that the
observed variation depends on both the benchmark problem and
the progression of systems within its sequences.
\autoref{fig:analysis-stored-trajectories} compares all five
methods on three complete sequences:
\textsc{Wheel}/QHW, \textsc{Murex}/IMCF, and \textsc{Pulley}/DSTV. The horizontal axis indexes recorded
matrices in the generation order $\nu=0,1,2\dots$.  The curves complement the aggregate statistics by
showing where iteration counts and solver costs change within
each sequence.

\paragraph{Time Breakdown}
\label{sec:runtime-breakdown}
To identify where ParaBMG spends its time, we inspect one
\textsc{Pulley}/LDSE system with 6.52 million scalar unknowns. At recorded matrix index 25, one numerical \emph{update} costs 16.38\,ms and the following ten-iteration PCG solve for the selected RHS costs 63.34\,ms. The combined cost is 79.72\,ms, of which \emph{update} and solve account for 20.5\% and 79.5\%, respectively (\autoref{fig:runtime-breakdown}, top).

A separate kernel profile of the same matrix and RHS shows
that block smoothing accounts for 44.6\% of kernel time,
outer PCG operations for 29.4\%, and multigrid residual
formation for 15.8\% (\autoref{fig:runtime-breakdown}, bottom). Transfers, precision conversion, and the terminal direct solve
account for the remaining approximately 10\%; the terminal solve
alone contributes 1.3\%. These shares are normalized by the
profile's 62.94\,ms of kernel activity, separately from the
unprofiled update-and-solve measurement.

The dominant work in this example lies in repeated smoothing
and matrix-vector operations. These are the operations targeted
by ParaBMG's decomposed block storage and iterate/residual
updates in the \emph{apply} stage
(\autoref{sec:decomposed-block-storage}), which avoid full
level-matrix products during smoothing and residual propagation.
The small terminal-solve share also shows that the prepared
coarsest solve is not the dominant cost in this instance.
The component comparisons that follow assess how the numerical
design and its execution contribute to performance.

\subsection{Additional Solver Comparisons}
\label{sec:analysis-direct}

The main benchmark compares iterative solvers designed to handle changing matrices. We additionally examine two alternatives: refactorizing each matrix with a sparse direct solver, and reusing a factorization of a fixed Poisson matrix as a preconditioner. These comparisons assess
the tradeoff between numerical update cost and the cost of the subsequent solve, as well as the memory required by factorization.

\paragraph{Setup.}
Direct linear solvers are less competitive when $N$ is large. We therefore sample the linear systems in different scales, and perform the experiments on \textsc{Blade}/IMCF and \textsc{Murex}/IMCF, with 176,545 and
1,647,843 unknowns, respectively. We sample the first, middle, and last linear system for each sequence.
The cuDSS runs on an A100 GPU with 80\,GB of memory. PARDISO and SuiteSparse CHOLMOD use eight CPU threads, while Eigen's native simplicial LDLT is serial. All direct solvers use double precision (FP64), perform symbolic
analysis once, and refactorize the numerical coefficients at each sampled state. PoissonPCG refers to use cuDSS to factorize a Poisson matrix once and reuses the resulting factorizations for every preconditioner \emph{apply} stage.
 \autoref{tab:analysis-direct} gives the timing details.

\begin{table}[tb]
  \centering
  \caption{\textbf{Additional solvers on sampled evolving systems.}   Mean update, solve, and total time in ms.   The last column is each method's $T$ divided by ParaBMG's $T$.   Initial preparation time is excluded. The OOM is short for out-of-memory, which reflects the failure during numeric factorization.}
  \label{tab:analysis-direct}
  \begin{tabular}{lrrrr}
    \toprule
    Solver & Update & Solve & Total & Speedup \\
    \midrule
    \multicolumn{5}{l}{\textsc{Blade}/IMCF ($N=176{,}545$)} \\
    cuDSS (GPU)       & 8.6    & 1.2   & 9.8    & $2.4\times$ \\
    PARDISO (CPU)     & 68.0   & 25.0  & 93.0   & $22.9\times$ \\
    CHOLMOD (CPU)     & 1070.7 & 43.4  & 1114.1 & $274.2\times$ \\
    Eigen LDLT (CPU)  & 1754.2 & 34.0  & 1788.2 & $440.2\times$ \\
    PoissonPCG (GPU)  & 0.0    & 54.5  & 54.5   & $13.4\times$ \\
    ParaBMG (GPU)     & 1.1    & 3.0   & \textbf{4.1} & $1.0\times$ \\
    \midrule
    \multicolumn{5}{l}{\textsc{Murex}/IMCF ($N=1{,}647{,}843$)} \\
    cuDSS (GPU)       & 71.4   & 6.7   & 78.1   & $3.5\times$ \\
    PARDISO (CPU)     & 833.8  & 289.3 & 1123.1 & $50.8\times$ \\
    CHOLMOD (CPU)     & >5,000 & --     & --      & -- \\
    Eigen LDLT (CPU)  & >5,000 & --     & --      & -- \\
    PoissonPCG (GPU)  & 0.0    & 530.5 & 530.5  & $24.0\times$ \\
    ParaBMG (GPU)     & 3.5    & 18.6  & \textbf{22.1} & $1.0\times$ \\
    \midrule
    \multicolumn{5}{l}{\textsc{Red Circular Box}/LDSE
    ($N=6{,}600{,}234$)} \\
    cuDSS (GPU)       & OOM     & --     & --      & -- \\
    ParaBMG (GPU)     & 30.7   & 71.5  & \textbf{102.1} & -- \\
    \bottomrule
  \end{tabular}
\end{table}

\paragraph{Results.}
ParaBMG achieves $2.4\times$ and $3.5\times$ speedups over cuDSS in update-plus-solve time on \textsc{Blade} and \textsc{Murex}, respectively, and $13.4\times$ and $24.0\times$ speedups over PoissonPCG (\autoref{tab:analysis-direct}). cuDSS has the lowest solve time, but the numerical factorization dominates its recurring cost. The cuDSS-based PoissonPCG avoids numerical updates by reusing fixed factorizations, but requires 42.0 and 75.7 iterations per linear system, compared with 9.3 and 14.3 for ParaBMG. Reusing fixed factorizations therefore reduces update costs at the expense of more expensive iterative solves as the systems evolve. On larger meshes with 6M unknowns, the direct solvers either exceed the available memory or fail to finish within the time limit.

\subsection{Full Pipeline Comparisons to References}

We conduct full pipeline comparisons with reference implementations of five geometry-processing applications: Dragon skinning weights for
QHW (Fig. 1 in \citeN{wang2021qhw}), Lucy mapping (\texttt{lucy\_o\_G}) for QHM (Fig. 1 in \citeN{wang2023qhm}), Pig local editing for LDSE (the Pig model beneath the table in Fig.1 of \citeN{chen2023ldef}), Sappho inflation using Iskra's inverse optimization of mean-curvature flow (Fig.3 in \citeN{dodik2026iskra}), and Nefertiti directional filtering using DSTV (M3; Fig. 1 in \citeN{brokman2024dstv}).
Dragon, Lucy, and Pig are the largest vertex-count examples in their
respective released test collections.

We retain the released meshes, constraints, and task settings for QHW,
QHM, and LDSE, and apply the released DSTV filtering settings to its
provided Nefertiti mesh. For IMCF, the reference code implementation uses an
unreleased local Sappho mesh file; we use the public Sappho scan from
Thingi10K (model 47984) \cite{zhou2016thingi10k}.

We compare the end-to-end runtime of our GPU pipelines using ParaBMG with the code implementation released by the reference publications. All experiments run on a Ryzen 7 9800X3D CPU and an RTX 5080 GPU, with an eight-thread CPU budget. The reference implementations use CPU TensorFlow and CHOLMOD for QHW, MATLAB for QHM and DSTV, the reference C++ implementation with oneMKL PARDISO for LDSE, and CPU PyTorch with Cholespy for IMCF.

\begin{table}[t]
\centering
\caption{End-to-end pipeline running time in seconds. Speedup is reference time divided by the time cost of our implementation. Both implementations execute the same task-defined budget in each case.}
\label{tab:reference-application-acceleration}
\begin{tabular}{lrrr}
\hline
Application & Reference & Ours & Speedup \\
\hline
QHW & 87.8 & 6.2 & $14.1\times$ \\
QHM & 29.4 & 7.5 & $3.9\times$ \\
LDSE & 6.3 & 1.7 & $3.7\times$ \\
IMCF & 130.9 & 3.3 & $39.7\times$ \\
DSTV & 47.8 & 10.8 & $4.4\times$ \\
\hline
\end{tabular}
\end{table}

Table~\ref{tab:reference-application-acceleration} reports end-to-end speedups of $14.1\times$, $3.9\times$, $3.7\times$, $39.7\times$, and $4.4\times$ for QHW, QHM, LDSE, IMCF, and DSTV, respectively, demonstrating the applicability of our approach across diverse geometry-processing workloads.

The reference implementations execute on the CPU using the C++, MATLAB, and CPU PyTorch backends described above, whereas our pipelines use the NVIDIA GeForce RTX 5080 GPU and incorporate ParaBMG. These pipeline-level speedups therefore jointly reflect GPU application kernels and linear solving. At identical tensor parameters, QHW and QHM agree with their reference solutions to relative errors of 7.0e-10 and 1.6e-9, respectively. LDSE's final vertex relative error is 7.1e-6, with no inverted tetrahedra and a $0.023\%$ difference in physical energy. IMCF's inflated and smoothed outputs agree with the reference to relative errors of 6.1e-5 and 2.9e-5; DSTV's filtered output agrees to 5.7e-12.

\section{Analysis}
\label{sec:solver-analysis}
We use ablation studies and comparative experiments to answer the following questions. (1) What does each of ParaBMG’s main components contribute to convergence and running time? (\autoref{sec:analysis-components})
(2) How well does ParaBMG scale to large linear systems as mesh resolution increases? (\autoref{sec:analysis-scalability})
(3) Why a simple combination of MAS and MG does not work directly, and how MAS and MG perform differently in elasticity simulation and geometry processing? (\autoref{sec:analysis-efficacy}) (4) As a boundary test, how robust is ParaBMG as the coefficient field becomes increasingly anisotropic along certain directions? (\autoref{sec:analysis-directional})

\subsection{Contributions of Interpolation, Block Smoothing, and Improved Execution}
\label{sec:analysis-components}

\begin{figure}[tb]
  \centering
  \includegraphics[width=\linewidth]{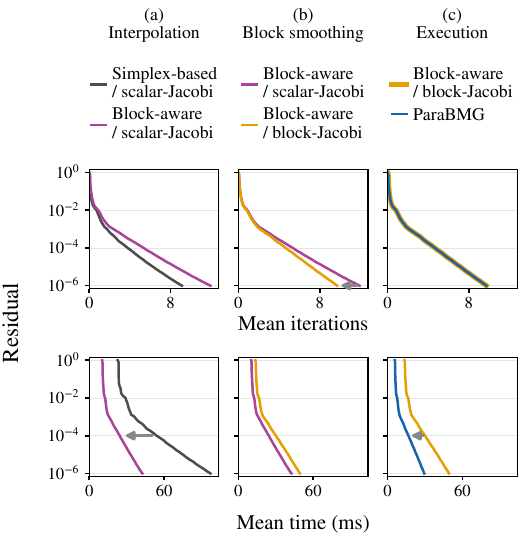}
  \caption{\textbf{Contributions of interpolation, block smoothing, and execution.}
  (a) Simplex-based and block-aware interpolation with scalar-Jacobi smoothing.
  (b) Scalar and block-Jacobi smoothing on the same block-aware hierarchy.
  (c) Optimized execution with the same hierarchy and block smoother.
  The upper row shows the mean iterations needed to first reach each
  residual. The lower row shows the
  corresponding mean time, including numerical update.
  Gray arrows indicate reduced time in (a) and (c), and fewer iterations
  in (b). The iteration curves in (c) coincide.}
  \label{fig:analysis-gravo-convergence}
\end{figure}

We examine three components through the paired comparisons in
\autoref{fig:analysis-gravo-convergence}: interpolation, block smoothing,
and execution. Each comparison uses the 
40 sequences of evolving linear systems.

\paragraph{Interpolation.}
We first compare Simplex-based / scalar with
Block-aware / scalar to examine the contribution of our hierarchy
with scalar-Jacobi smoothing. This change reduces the combined update and
apply time by factors of $1.3\times$ on triangle meshes and $2.1\times$
on tetrahedral meshes. The hierarchy therefore provides a time benefit
compared to the GravoMG's simplex-based interpolation, even
before block smoothing is introduced.

\paragraph{Block Smoothing.}
We next compare Block-aware / scalar with Block-aware / block,
keeping the hierarchy and full-CSR execution path fixed.
Block-Jacobi reduces iteration counts by factors of 1.24 and 1.21
on triangle and tetrahedral meshes, respectively.
However, the additional cost of constructing and applying block
inverses outweighs this convergence benefit in the tested CSR
implementation, increasing the combined update and solve time.
This motivates adapting the execution to the block smoother.

\paragraph{Improved Execution.}
Finally, we retain the same hierarchy, block partition, and smoothing
parameters, and switch to ParaBMG's execution based on the
block-diagonal matrix $\bD$  and the off-diagonal cross-block $\bE$ decomposition.
Precompiled contribution maps accelerate coarse matrix updates,
while fused smoothing and residual vector updates avoid
redundant matrix-vector products.
The iteration counts therefore remain unchanged, while the combined
update and solving time decreases by factors of $1.5\times$ on triangle
meshes and $1.7\times$ on tetrahedral meshes.  
With these execution optimizations, ParaBMG also outperforms the
  Block-aware / scalar configuration on the same hierarchy by factors
  of $1.38\times$ on triangle meshes and $1.44\times$ on tetrahedral
  meshes. This reverses the cost ranking observed with the full-CSR
  implementation, where the convergence benefit of block Jacobi translates
  into lower total cost. A similar optimization strategy, such as applying a precompiled contribution maps
  for the CSR-based scalar-Jacobi, does not result in a comparable performance gain.
These results demonstrate how matching the matrix representation
and execution to the block smoother translates improved convergence
into lower running time.

\textbf{Remark}   The ``Simplex-based / scalar-Jacobi'' configuration used as the starting  point of this ablation differs from the GravoMG-GPU baselines in  \autoref{tab:evaluation-applications} and already outperforms them.   It incorporates an outer conjugate gradient loop for more robust  convergence and replaces the original smoothers with scalar Jacobi.  This avoids the sparse triangular solves required by Gauss-Seidel smoothing in the surface method~\cite{wiersma2023gravo}, as well as
  the time-consuming spectral estimation by a fixed number of power iterations in the volumetric method~\cite{padilla2026gravo}.
\subsection{Scalability with Mesh Resolution}
\label{sec:analysis-scalability}

To examine scalability as mesh resolution increases, we solve
$-\Delta u=f$ on seven tetrahedralizations of \textsc{Murex},
a mesh with fine geometric detail from our ten mesh dataset.
The meshes range from 1,276 to 2,418,269 vertices, with finer
resolutions capturing more surface detail.
We use linear tetrahedral finite elements with homogeneous Dirichlet
conditions on the exterior boundary.

We compare ParaBMG with all considered baselines including iterative and direct solvers.
Iterative methods start from zero and use a relative residual of
$1\mathrm{e}{-6}$.
We measure the mean numerical update and apply time, excluding
initial setup. Full settings and results are provided
in \autoref{tab:analysis-resolution}.

\autoref{tab:analysis-resolution} reports speedups relative to the
fastest tested baseline at each resolution.
For the two meshes with fewer than 5,000 vertices, no multilevel hierarchy is constructed, and ParaBMG reduces to a single level with a direct solve.
At 13K and 50K vertices, cuDSS is the strongest tested baseline,
and ParaBMG achieves speedups of $1.5\times$ and $3.9\times$,
respectively. At 190K, 758K, and 2.4M vertices, BoomerAMG is the
strongest tested baseline, with corresponding ParaBMG speedups
of $5.7\times$, $3.9\times$, and $3.5\times$.

Across the five resolutions that use multilevel correction,
the mean iteration count grows modestly from 8.0 to 13.7.
At 2.4M vertices, ParaBMG requires 39.2\,ms for numerical update
and apply. These results show that ParaBMG maintains effective
convergence as resolution increases and retains a runtime advantage
over the tested baselines on the larger meshes.

\begin{table}[t]
  \centering
  \caption{\textbf{Scalability on the Murex volume.}
  $N$ is the number of vertices, $\overline{n}_{\text{cg}}$ is the mean
  PCG iteration count, and $T$ is the mean numerical update and apply
  time in milliseconds. \emph{Acc} refers to the relative acceleration ratio with respect to the fastest tested
  baseline at each reported resolution.
  The inset shows the coarsest volume mesh, finest volume mesh, and the original mesh.
  Both volumes use a common cutting plane and physical color scale
  for the solution of $-\Delta \mathbf{u}=\mathbf{1}$ with homogeneous Dirichlet conditions.}
  \label{tab:analysis-resolution}
  \label{fig:murex-volume-cutaway}
  \begin{minipage}[c]{0.48\linewidth}
    \centering
    \setlength{\tabcolsep}{4pt}
    \begin{tabular}{rrrr}
      \toprule
      $N$ & $\overline{n}_{\text{cg}}$ & $T$ & Acc. \\
      \midrule
      1.2K   & -- & 0.7  & -- \\
      3.7K   & -- & 1.2  & -- \\
      13K    & 8.0  & 2.1  & $1.5\times$ \\
      50K    & 9.2  & 3.4  & $3.9\times$ \\
      190K   & 11.0 & 5.4  & $5.7\times$ \\
      758K   & 12.0 & 14.3 & $3.9\times$ \\
      2.4M & 13.7 & 39.2 & $3.5\times$ \\
      \bottomrule
    \end{tabular}
  \end{minipage}\hfill
  \begin{minipage}[c]{0.50\linewidth}
    \centering
    \includegraphics[width=\linewidth]{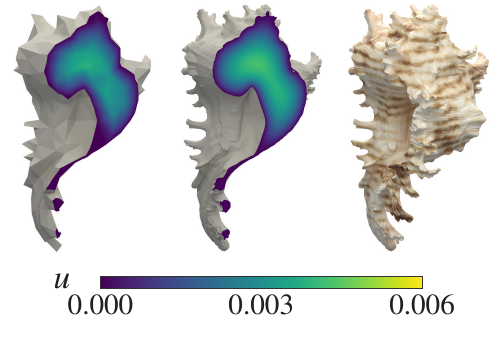}
  \end{minipage}
\end{table}

\subsection{Rationale and Challenges of MAS-MG Combinations}
\label{sec:analysis-efficacy}
\label{sec:analysis-mas-mg}

This section examines the local-global error reduction motivation underlying ParaBMG and the design choices that distinguish it from earlier methods. We first test whether coarse-grid correction and parallel block solves complement each other by addressing errors at different spatial scales.  Two experiments support this motivation: incorporating MG corrections into a MAS framework substantially reduces the iteration count, while  increasing the block-Jacobi size in AMGX reduces PCG iterations by a margin comparable to that observed in \autoref{tab:evaluation-applications}.   However, these convergence improvements do not automatically translate into comparable performance gains. Effective integration requires suitable  algorithmic choices together with efficient GPU execution.
 
  We then examine why related methods have emphasized different design
  choices. CPU-oriented geometric multigrid methods \cite{liu2021surface,wiersma2023gravo} use smoothers well suited to serial  execution, where the tradeoffs differ from those motivating parallel  block solves and GPU-oriented storage. For MAS, a separate elasticity diagnostic examines how its effectiveness depends on the operator and  workload, explaining why solver rankings in physics simulation need not  carry over to geometry processing.
  
\paragraph{Multigrid Composition of Local Corrections.}
Multilevel additive Schwarz (MAS) combines local corrections across
levels and has been used in deformable body
simulation~\cite{wu2022mas,huang2023gipc,huang2024stiffgipc,wang2026agipc}.
We examine whether multigrid composition can improve its performance.  Our experimental MAS+MG variant retains the MAS hierarchy,
  local block solvers, and factorizations, but replaces the additive
  combination of level corrections with a symmetric multigrid V-cycle.
  At each level, one damped block-Jacobi pre-smoothing
  step is followed by restriction of the updated residual, a recursive
  coarse correction, and one post-smoothing step on the resulting
  residual. We retain the original MAS-style piecewise-constant prolongation and use its transpose for restriction. 

We compare the methods on the complete \textsc{Wheel}/QHW sequence
of 301 systems with 784,680 unknowns. 
MAS+MG reduces the mean iteration count from 253.90 to 135.26 and
the mean update and solve time from 539.85 to 407.86\,ms.
The reduction in iterations offsets the additional work per
iteration, yielding a $1.3\times$ speedup.

\begin{figure}[tb]
  \centering
  \includegraphics[width=\columnwidth]{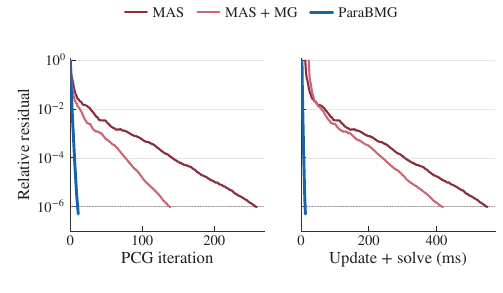}
  \caption{\textbf{Multigrid composition improves the tested MAS preconditioner.}
  Residual versus PCG iteration and the combined update and solve
  time for one system from \textsc{Wheel}/QHW.
  The text reports means over the complete sequence.}
  \label{fig:analysis-mas-comparison}
\end{figure}

ParaBMG requires 10.22 iterations and 12.08\,ms on average over
the same sequence. Multigrid composition improves the tested MAS
preconditioner, while MAS+MG's mean cost remains $33.8\times$
that of ParaBMG. This remaining gap motivates the following
comparisons of interpolation, block smoothing, and execution
to examine how these choices contribute to ParaBMG's performance.

\paragraph{Block-Jacobi in Multigrid.}
To isolate the effect of local solve size, we vary the block Jacobi
size within the AMGX implementation on one typical Pulley/LDSE
system used in \autoref{sec:runtime-breakdown}. The $q=1$ setting uses scalar-Jacobi. Larger values of $q$
allow each local solve to jointly correct more coupled unknowns.
All other hierarchy and relaxation settings remain fixed.
For $q=1,2,4,8,16,32$, the PCG iteration counts are
35, 35, 34, 32, 31, and 29, respectively
(\autoref{fig:analysis-amgx-block-size}). Increasing $q$ from 1 to
32 reduces the iteration count by 17.1\%, while the solve time
increases from 255.9 to 420.2\,ms.

\begin{figure}[tb]
  \centering
  \includegraphics[width=\columnwidth]{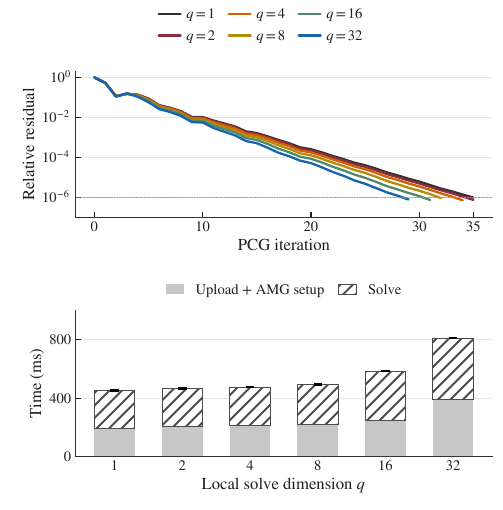}
  \caption{\textbf{Larger local solves trade fewer iterations for additional work.}
  Local solve sizes from $q=1$ to $q=32$ within a fixed AMGX hierarchy.
  Top: residual versus PCG iteration.
  Bottom: matrix upload and full AMG setup (gray), followed by solve
  (hatched). Bars show mean times over three independent preparations.
  Error bars show the sample standard deviation of the total time.
  The timing includes initial preparation as well as the solve.}
  \label{fig:analysis-amgx-block-size}
\end{figure}

Larger local solves improve convergence on this hierarchy, although
their additional work increases the solve time. For ParaBMG, this
motivates reducing the cost of block smoothing. In our ParaBMG setting,
we choose the block capacity $m_{\text{agg}}=32$
that preserves the convergence benefit while keeping the local solve cost low. This also suggests that an efficient block-Jacobi smoothing in this case requires substantial algorithmic designs and execution optimizations discussed in \autoref{sec:analysis-components}.

\paragraph{Mass Screening in Vector Elasticity.}
To identify where simulation-oriented solvers such as MAS begin to
degrade as the linear systems move toward those of geometry
processing, we relate the two settings through mass screening.
Each implicit simulation step solves a system
$\bK+\Delta t^{-2}\bM$, the vector-valued analogue of a screened
Poisson equation, in which the mass term acts as the screening term.
At interactive time steps, this screening term is large, so errors
remain largely local and inexpensive local block
solves are highly effective.
Geometry processing pipelines, in contrast, typically produce weakly
screened or unscreened Poisson-type systems whose coupling extends
across the entire mesh, so mesh-wide low-frequency errors dominate and
coarse-grid correction becomes essential.
We therefore vary the screening strength on a single elasticity
problem, moving continuously from simulation-like to
geometry-processing-like systems.
We test a hanging Armadillo inspired by
AGIPC's multigrid comparison~\cite[Sec.~5.1]{wang2026agipc}, using a
Neo-Hookean material ($E=0.1$\,MPa,
$\nu=0.35$, $\rho=10^3$\,kg/$m^3$). The  system has 94,458 scalar unknowns. With an upper-back attachment, backward Euler
integrates 1.5\,s under gravity at $\Delta t=1/60$\,s, without contact.  
MAS uses the StiffGIPC implementation while the AMGX uses FP64 $3\times3$ BSR,
\texttt{AGGREGATION}/\allowbreak\texttt{SIZE\_4}, symmetric
\texttt{MULTICOLOR\_GS}, and one or ten V-cycles per application,
following the configuration study in AGIPC's supplemental
Sec.~4~\cite{wang2026agipc}. Here, ParaBMG is not specially tailored to $3N\times 3N$ systems. Instead, it operates directly on an expanded mesh graph, with three nodes per mesh vertex and full edge connections both within each triple and between triples corresponding to adjacent mesh vertices.
 Table~\ref{tab:analysis-armadillo-replay}
separately reports the deformed Newton systems.

\begin{figure}[tb]
  \centering
  \includegraphics[width=\columnwidth]{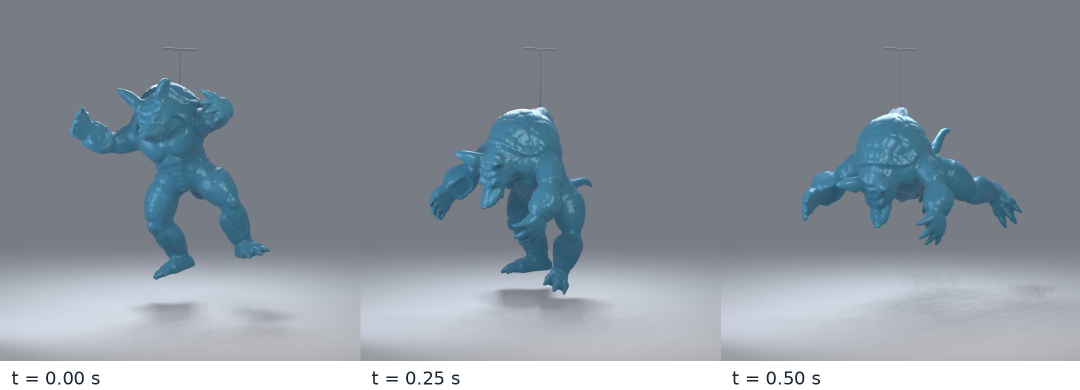}
  \caption{\textbf{Armadillo elasticity.}
  Physical states illustration at 0, 0.25, and 0.5\,s.}
  \label{fig:analysis-armadillo}
\end{figure}

We also freeze the rest-state stiffness and vary only the mass coefficient
  $0\leq\alpha\leq10^{10}\;\mathrm{s}^{-2}$ defined in $\bA(\alpha)=\bK_{\mathrm{rest}}+\alpha\bM$.
Here $\bM$ includes density and the actual time step gives
$\alpha=\Delta t^{-2}=3600$\,s$^{-2}$. The tuning of $\alpha$ controls the difficulty of the problems.
\begin{figure}[tb]
  \centering
  \includegraphics[width=\linewidth]{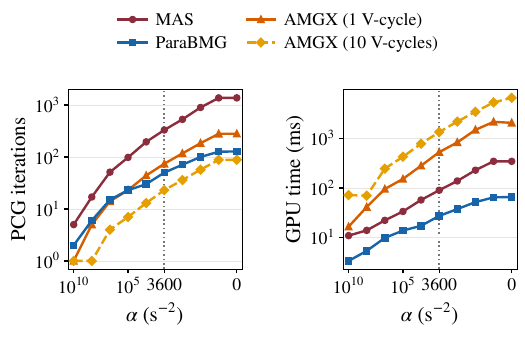}
 \caption{\textbf{Changing the problem difficulties on one vector-elasticity system.}
  All  use the same Armadillo geometry, stiffness, constraints, and
  gravity RHS, with relative residual tolerance $10^{-6}$. Left: PCG
  iterations. Right: numerical-update-plus-PCG GPU time. Samples of $\alpha$
  are ordered from strong screening to the static endpoint. The dotted
  line marks the actual time step, $\alpha=3600$\,s$^{-2}$.}
  \label{fig:analysis-elastic-screening}
\end{figure}
As screening weakens, the systems transition from dynamics-like problems toward the static elasticity limit, and both iteration counts and computational costs generally increase for all methods. MAS consistently requires more iterations than the AMGX configurations, yet achieves lower GPU update-plus-solve time, demonstrating that iteration count alone does not determine efficiency. ParaBMG requires fewer iterations and achieves lower update-plus-solve time than MAS.
Reducing $\alpha$ weakens the mass term while preserving the vector couplings and deformation modes of the elasticity operator. Within this family, MAS retains its timing advantage over AMGX even at the static endpoint. On the Poisson-like systems in our geometry-processing benchmarks, MAS incurs a substantially larger iteration penalty relative to multigrid. The cost of these additional iterations can outweigh savings in updates or individual iterations.

\paragraph{Scope and Implications.} This experiment shows that mass screening shifts the balance between iteration count and per-iteration cost. As screening weakens toward the unscreened limit characteristic of geometry processing, MAS's iteration penalty relative to multigrid grows faster than its per-iteration savings, widening its gap to ParaBMG. The operator also matters: MAS remains faster than AMGX across this elasticity family, and even its static-elasticity gap to ParaBMG is smaller than on our scalar Poisson benchmarks.

StiffGIPC improves the MAS aggregation and GPU operations on the block-sparse Hessian~\cite{huang2024stiffgipc}, and supports specified dynamic structural changes induced by contact. The elasticity test therefore serves as a diagnostic experiment. Designing a solver tailored to elasticity with evolving contact would require coarse spaces that capture low-energy vector modes, GPU storage and kernels that exploit the dense $3\times3$ couplings, and efficient handling of contact-induced changes in sparsity. Jointly designing these components lies beyond the scope of this work.

\begin{table}[tb]
  \centering
  \caption{\textbf{Armadillo Newton-system replay at $10^{-6}$.}
  Mean PCG iterations and GPU
  update-plus-solve time $T$ are averaged over successful systems only.}
  \label{tab:analysis-armadillo-replay}
  \begin{tabular}{lrrr}
    \toprule
    Method & $\bar n_{\mathrm{cg}}$ & $T$ (ms) \\
    \midrule
    MAS & 318.9 & 86.7 \\
    ParaBMG & 50.4 & 27.1\\
    AMGX (1 V-cycle) & 71.6 & 449.2 \\
    AMGX (10 V-cycles) &  21.9 & 1317.3 \\
    \bottomrule
  \end{tabular}
\end{table}

\subsection{Robustness under Directional Coupling}
\label{sec:analysis-directional}

To test the limits of geometric coarsening under strongly directional
coupling, we construct an anisotropic Poisson system on a
$48\times12\times12$ grid subdivided into 41,472 tetrahedra with
8,281 nodes and a cell-edge aspect ratio of four. Piecewise-linear
tetrahedral finite elements discretize
$-\nabla\!\cdot(\mathbf K\nabla u)=f$, with
\begin{equation}
  \mathbf K=\mathbf I+(\kappa-1)\mathbf d\mathbf d^{\mathsf T},
  \qquad \kappa=10^5,\quad \mathbf d=(1,0,0)^{\mathsf T}.
  \label{eq:analysis-directional-tensor}
\end{equation}
This tensor deliberately couples the $x$ direction much more strongly
than the other two directions.
Dirichlet conditions apply on the two $y$ faces, with natural conditions
elsewhere.

We compare geometric coarse vertex selection strategy (PMIS-only) with an experimental
coefficient-aware alternative, AHAT-D2. These two coarse vertex selection strategies are introduced in \autoref{sec:method_graph}, which generate a geometry-aware hierarchy and a coefficient-aware hierarchy, respectively. The geometry-aware hierarchy fails to reach a relative residual of $10^{-6}$ within 500 PCG iterations, whereas the coefficient-aware hierarchy converges in eight iterations with relative residual of   $4.81\times10^{-7}$.
  This result is expected in this extreme setting, as a purely geometric
  hierarchy does not adapt its coarsening and interpolation to the
  coefficient distribution and therefore cannot be expected to remain
  effective for all cases especially involving extreme coefficient variation. It suggests that a coefficient-aware hierarchy rebuilding can be beneficial when coefficients of the matrix become highly nonuniform.  However, we observe no comparable degradation in the geometry processing
  benchmarks in \autoref{tab:evaluation-applications} or the interactive
  applications in \autoref{sec:applications}.
  We therefore only report the performance of the geometry-aware hierarchy performance in \autoref{tab:evaluation-applications} and \autoref{sec:applications}, and retain the coefficient-aware hierarchy rebuilding as an optional, user-controlled strategy for extreme cases.

\begin{figure*}[t]
\centering
\includegraphics[width=\linewidth,height=.225\textheight,keepaspectratio]{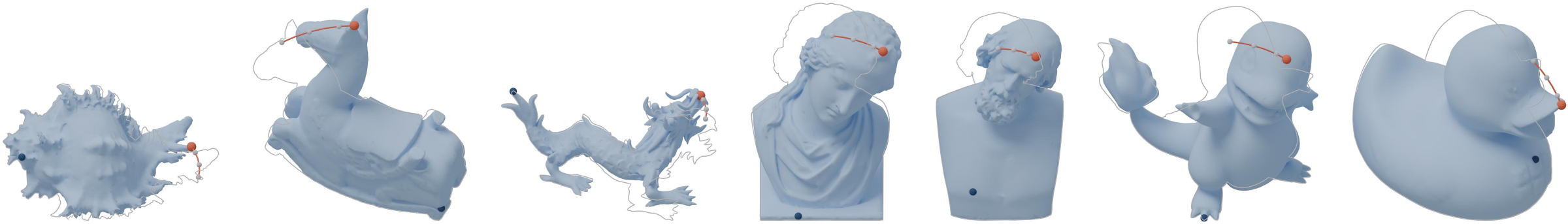}
\caption{Mesh editing (LDSE) on the seven-model joint scene, rendered in Blender.
Five control updates produce $30^\circ$ terminal rotations.
Gray contours show initial silhouettes. Navy markers indicate anchors,
while terracotta markers show the moved controls.}
\label{fig:application-editing}
\end{figure*}

\section{Applications}
\label{sec:applications}

We demonstrate interactive applications of ParaBMG in three established
geometry processing pipelines: local deformation for shape editing
(LDSE), variational quasi-harmonic mapping (QHM), and quasi-harmonic
interpolation weights (QHW). Each scene contains the seven teaser
models: Murex, Isidore Horse, XYZRGB Dragon, Aphrodite, Tennyson,
Charmander, and Duck. Each interaction updates all seven models.
The figures show results rendered in Blender, while all timings
are measured in the interactive WebGL application. We refer readers to the supplementary document for details of the settings.

We compare ParaBMG and AMGX within the same GPU application pipelines,
using the integrated reference implementations as CPU baselines.
We measure backend update latency from receipt of an interaction
request to preparation of the numerical results, excluding one-time
initialization and rendering. Reported latencies are medians over
three runs. All tests are conducted on a workstation with an
RTX 5080 GPU and a Ryzen 7 9800X3D CPU, using eight CPU threads.

Across all three applications, ParaBMG reduces backend update latency
while maintaining numerical agreement.
For LDSE, QHM, and QHW, respectively, ParaBMG achieves speedups of $1.34\times$, $1.58\times$, and $2.86\times$ over the strongest baseline backend solver AMGX, and $15.15\times$, $6.48\times$, and $184.56\times$ over the CPU reference implementations.
These improvements bring texturing and weight updates below one
second and enable approximately one-second coarse previews for
mesh editing.

\begin{figure*}[]
\centering
\includegraphics[width=\linewidth,height=.225\textheight,keepaspectratio]{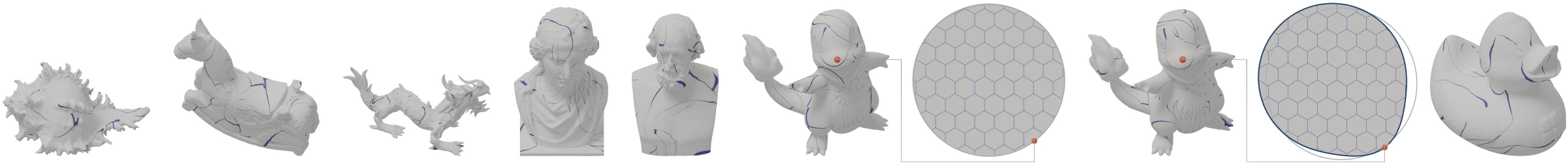}
\caption{Interactive texture placement using QHM on seven surface meshes.
  Each surface is cut along mesh edges into a disk and mapped to a planar UV domain.
  Dragging boundary control points updates the UV mapping while keeping
  the hexagonal texture and 3D geometry fixed.
  The planar domains and adjacent surface views illustrate the mapping
  before and after a boundary edit.
  The blue-purple curves on the surfaces are the mapped edges of the
  hexagonal texture.}
\label{fig:application-texturing}
\end{figure*}

\paragraph{Mesh Editing}
Following \citet{chen2023ldef}, we combine Neo-Hookean elasticity
with a smoothly clamped $\ell_1$ displacement regularizer.
The ADMM optimization alternates local updates with global position
solves. The global position step solves three $N\times N$ linear equations, which we accelerate using ParaBMG. The dynamic re-weighting is deployed to produce different systems per optimization steps.
We use the same initialization and a fixed budget of 50 ADMM
iterations to produce coarse previews.
\autoref{fig:application-editing} shows the results after five
control updates, producing $30^\circ$ terminal-region rotations
on seven objects.
For a representative single $12^\circ$ control edit, ParaBMG
completes the backend update in approximately one second.
Including communication and display updates, the application
achieves 1.07 updates/s, providing the feedback with interactive previews
across all seven models.

\paragraph{Texturing}
Following \citet{wang2023qhm}, we update texture placement by
optimizing anisotropic surface tensors and solving for two
texture-coordinate fields.
Changes to the tensors produce evolving linear systems with
fixed connectivity. Users drag the boundary of a planar domain to adjust texture
placement while the 3D geometry remains fixed.
\autoref{fig:application-texturing} shows three boundary drags,
with five optimization steps per update.
Including communication and display updates, the application
achieves 1.58 updates/s, enabling repeated boundary adjustments
with visual feedback across all seven shapes.

\begin{figure*}[]
\centering
\includegraphics[width=\linewidth,height=.225\textheight,keepaspectratio]{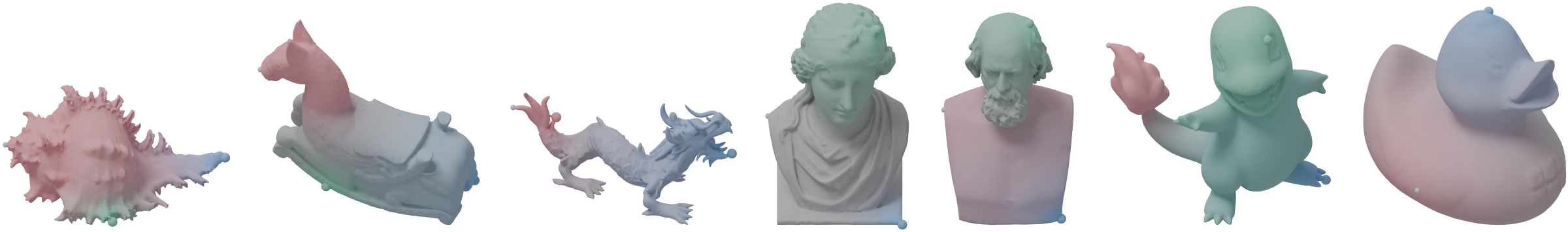}
\caption{Interpolation weights (QHW) on the seven-model joint scene, rendered in Blender. Colored spheres identify three influences on each object. Segments
join initial and updated handles. Surface colors visualize the resulting blended weight fields.}
\label{fig:application-weighting}
\end{figure*}

\paragraph{Interpolation Weights}
Following \citet{wang2021qhw}, we optimize anisotropic tensor
fields to compute interpolation weights for three control groups
on each object.
Moving the control handles triggers optimization and linear
solves to update the weight fields.
\autoref{fig:application-weighting} shows the resulting fields
after handle relocation.
With three optimization steps and a final evaluation, ParaBMG
completes each backend update in 162\,ms.
Including communication and display updates, the application
achieves 4.66 updates/s, enabling repeated handle adjustments
with visual feedback.

\paragraph{Application vs. Solver Speedups}
Each interactive edit triggers a nonlinear optimization loop
  comprising a sequence of linear solves and additional computations.
  These additional computations are common to the AMGX backend and the
  ParaBMG backend, limiting the reduction in application
  response time achieved by replacing the linear solver.
  Consequently, the application-level speedups are smaller than
  those reported in the solver benchmarks in
  \autoref{tab:evaluation-applications}.
Nevertheless, with the ParaBMG backend and our accelerated implementation of the whole pipelines, these applications remain substantially
faster than the original CPU reference implementations.
\section{Conclusion, Limitation, and Future Work}

ParaBMG provides an effective framework for large-scale, evolving Poisson systems, achieving speedups of $3.7\times$ to $7.7\times$
  in average solver cost over the fastest tested GPU baseline for each
  of 40 sequences comprising 40,987 instances. The framework combines
  fully parallel geometric hierarchy construction and reuse with block
  Jacobi smoothing and efficient GPU operators for repeated solves.
  Our three-stage ablation studies show that ParaBMG's hierarchy improves the performance even with scalar smoothing, block smoothing further reduces iteration counts, and the optimized parallel implementation preserves these counts and reduces the solver cost. Separate comparisons on the reference examples demonstrate acceleration of complete GPU pipelines over the released CPU implementations under matched workloads. These gains reflect the combined effects of GPU geometry processing and linear solving.  Beyond these reference comparisons, ParaBMG enables three interactive   geometry processing applications on high resolution meshes: mesh  editing, surface parameterization, and interpolation weight computation. These applications achieve speedups up to orders of magnitude over their original CPU implementations, supporting interactive geometry processing applications.

Our method is not without limitations. Automated tuning of the solver hyperparameters is a promising direction that is underexplored for
further improving performance across different meshes and coefficient
distributions, including but not limited to the coefficient-aware rebuilding strategy when facing extreme cases with strong coefficient variation.
Another direction is to broaden the framework beyond its current
mesh-based SPD Poisson setting.  More general stiffness-matrix problems, including coupled
$3N \times 3N$ systems with three unknowns per mesh vertex, simultaneous
solving of multiple right-hand-side vectors, and non-SPD solves are also
important topics not addressed in this paper. These extensions offer
opportunities to incorporate coupled-variable coarse representations,
exploit shared work across right-hand sides, and integrate iterative
methods suited to broader operator classes. Investigating them would
extend ParaBMG's applicability while building on the parallel
construction and efficient repeated-solve mechanisms established here.

\bibliographystyle{ACM-Reference-Format}
\bibliography{reference}

\appendix
\section{Details of Benchmark Problems}
\label{supp:benchmarks}

We describe the five problem formulations and the settings used to
generate their evolving linear systems. LDSE, QHW, IMCF, and DSTV
provide the four problem families in the current main benchmark. We choose them since they accept watertight mesh inputs, while the rest QHM is typically applied on disk-like surface meshes, and therefore excluded from the main benchmarking dataset. For the interactive applications, we instead consider the QHW, QHM and LDSE, which supports interactive demonstrations.
The application demonstrations may use different meshes and interaction
sequences. Their settings are distinguished from the benchmark data.

\paragraph{Shared notation and discretization.}
We retain $\mathcal M=(V,E,\mathcal T)$, $N=|V|$, $d$,
$\mathbf G_\tau$, $\mathbf C_\tau^{(\nu)}$, and
$\mathbf M^{(\nu)}$ from the main paper. For convenience, define the
stiffness contribution
\begin{equation}
 \mathbf K^{(\nu)}
 =\sum_{\tau\in\mathcal T}|\tau|\,
    \mathbf G_\tau^\top\mathbf C_\tau^{(\nu)}\mathbf G_\tau,
 \qquad
 \mathbf A^{(\nu)}=\mathbf K^{(\nu)}+\mathbf M^{(\nu)}.
\end{equation}

\paragraph{Constraints and normalization.}
For free indices $F$ and prescribed indices $D$, the benchmark retains
the full dimension using symmetric Dirichlet treatment:
\begin{align}
 \mathbf A_{FF}&=\widetilde{\mathbf A}_{FF},&
 \mathbf A_{DD}&=\mathbf I,&
 \mathbf A_{FD}&=\mathbf A_{DF}=\mathbf0,\\
 \mathbf b_F&=\widetilde{\mathbf b}_F-
               \widetilde{\mathbf A}_{FD}\mathbf x_D,&
 \mathbf b_D&=\mathbf x_D.
\end{align}
Here tildes denote the assembled quantities before constraints.
The benchmark scales the free equations by the median free diagonal
and normalizes each nonzero RHS by its Euclidean norm, retaining the
normalization needed to recover physical solution values.
All compared methods receive the same saved equations. For the interactive applications, we incorporate the ParaBMG into the pipelines to support the online solving.

\begin{table}[t]
\centering\small
\caption{Number of linear systems for each benchmark mesh and problem.
All counts and totals are for one method.}
\begin{tabular}{@{}lrrrrr@{}}
\toprule
Mesh & LDSE & QHW & IMCF & DSTV & Total \\
\midrule
Wheel & 150 & 2,408 & 1,290 & 294 & 4,142 \\
Blade & 150 & 2,408 & 1,140 & 243 & 3,941 \\
Vase Lion & 150 & 2,408 & 1,404 & 243 & 4,205 \\
Duck & 150 & 2,408 & 1,398 & 348 & 4,304 \\
Rocker Arm & 150 & 2,408 & 1,050 & 338 & 3,946 \\
Murex & 150 & 2,408 & 1,728 & 310 & 4,596 \\
Isidore Horse & 150 & 2,408 & 1,290 & 377 & 4,225 \\
Pulley & 150 & 2,408 & 978 & 398 & 3,934 \\
Oil Pump & 150 & 2,408 & 1,086 & 313 & 3,957 \\
Red Circular Box & 150 & 2,408 & 1,014 & 165 & 3,737 \\
\midrule
Total & 1,500 & 24,080 & 12,378 & 3,029 & 40,987 \\
\bottomrule
\end{tabular}
\end{table}

\textbf{Remark} During the benchmarking, the input initial guess vectors for the iterative solvers are from the actual records. For example, for one x-direction right-hand-side vector instance for the LDSE problem, its initial guess vector should be the x-directional vector at the last solve.
In practice, for the application problems like LDSE that solve 3 right-hand-side vectors together, one could call the \emph{update} stage only once for these three solves, or develop more efficient solvers that solves the condensed system collectively. However, developing such pipelines requires substantial changes and careful optimization in the implementation, and we leave them to future work. To obtain a fair comparison, we make sure that the solving time always contain one ${update}$ stage and one PCG solve.

\subsection{Local Deformation}

Local deformation for interactive shape editing
(LDSE)~\cite{chen2023ldef} uses local-global optimization on a
tetrahedral mesh. The local steps update elastic targets and
displacement targets. The global step solves three coordinate
columns sharing a scalar SPD matrix. This is distinct from solving
a coupled $3N\times3N$ elasticity matrix.

\paragraph{Linear systems.}
Let $\eta_\tau^{(\nu)}>0$ denote the elastic penalty including
the element weight, and let $\eta_0>0$ be the displacement penalty.
Before imposing prescribed positions, the global step has
\begin{align}
 \widetilde{\mathbf A}^{(\nu)}
 &=\sum_\tau\eta_\tau^{(\nu)}
                   \mathbf G_\tau^\top\mathbf G_\tau+\eta_0\mathbf I,\\
 \widetilde{\mathbf B}^{(\nu)}
 &=\sum_\tau\eta_\tau^{(\nu)}\mathbf G_\tau^\top
             (\mathbf Z_\tau^{(\nu)}-\mathbf U_\tau^{(\nu)})
        +\eta_0\mathbf T^{(\nu)}.
\end{align}
The matrices $\mathbf Z_\tau^{(\nu)}$ and $\mathbf U_\tau^{(\nu)}$
are the local elastic target and multiplier, and
$\mathbf T^{(\nu)}$ contains the displacement targets.
In the notation of the main paper,
$\mathbf C_\tau^{(\nu)}
=(\eta_\tau^{(\nu)}/|\tau|)\mathbf I$ and
$\mathbf M^{(\nu)}=\eta_0\mathbf I$.
Reference gradients remain fixed while adaptive penalties change the
matrix coefficients. The positive displacement term makes the global
quadratic SPD even though the outer deformation objective is nonconvex.

\paragraph{Dataset generation.}
The dataset generator enables adaptive element reweighting with zero
delay and an aggressive weight-ratio threshold of one. Each mesh supplies ten independent edits, each starting from the rest shape. An edit pins a
root region and moves a tip region outward by 10\% of the bounding-box
extent. Five steps are performed per edit, producing
50 matrices and 150 linear system solving per mesh in the current benchmark. The initial guess is reset at the beginning of each edit.
Within an edit, each coordinate can reuse the preceding valid solution
for that coordinate.

\paragraph{Application settings.} We support an interactive WebGL-based application for the LDSE pipeline.
For the timed LDSE update, we choose each object's longest bounding-box axis on its exterior vertices. Vertices below its 12th percentile are fixed, and those above its 97th percentile form the grabbed region. We rotate that region
by $12^\circ$ about the exterior centroid in the plane of this axis and the
cyclically next Cartesian axis. All backends receive identical initial
coordinates obtained by smoothly interpolating the fitted rigid control
motion and imposing the prescribed positions exactly; this preparation is
timed. We execute 50 ADMM iterations with stopping tolerance $10^{-30}$ and
adaptive element reweighting disabled, unlike the benchmark generator above.
The Neo-Hookean model uses Young's modulus $10^7$, Poisson ratio $0.45$,
displacement penalty $\eta_0=10^3$, and smoothly clamped $\ell_1$ weight $10^4$
with threshold $0.1$. Local updates allow 20 iterations: damped Newton on the
GPU and the reference local L-BFGS implementation on the CPU. The original
rest mesh is retained between interactions. Each global step has three
coordinate RHS columns sharing one scalar matrix. The illustration and video
use a separate sequence of five $6^\circ$ commits, reaching the same
$30^\circ$ rotation on every object.
\subsection{Quasi-Harmonic Maps}

Quasi-harmonic maps (QHM)~\cite{wang2023qhm} optimize a diffusion
tensor on each triangle and solve for two planar coordinates.
For fixed tensors and prescribed boundary coordinates, the energy is
\begin{equation}
 \frac12\sum_\tau|\tau|\,
 \operatorname{tr}\!\left[
  (\mathbf G_\tau\mathbf Y)^\top
  \mathbf C_\tau^{(\nu)}(\mathbf G_\tau\mathbf Y)\right],
 \qquad \mathbf Y=[\mathbf u,\mathbf v].
\end{equation}
The free RHS entries arise from coupling to the prescribed boundary.

\paragraph{Tensor parameterization.}
For a two-component parameter $\boldsymbol\theta$ with
$\|\boldsymbol\theta\|_2<1$, the implementation uses
\begin{equation}
 \mathbf C(\boldsymbol\theta)=
 \frac{1}{1-\theta_1^2-\theta_2^2}
 \begin{bmatrix}
 (1-\theta_1)^2+\theta_2^2&-2\theta_2\\
 -2\theta_2&(1+\theta_1)^2+\theta_2^2
 \end{bmatrix}.
\end{equation}
Its determinant is one and its eigenvalues are
$(1+\|\boldsymbol\theta\|_2)/(1-\|\boldsymbol\theta\|_2)$ and
its reciprocal.
A radial parameterization
$\boldsymbol\theta=\tanh(\|\boldsymbol\xi\|_2)
\boldsymbol\xi/\|\boldsymbol\xi\|_2$, with the continuous value at
zero, enforces the admissible disk in exact arithmetic.
Updating the tensors changes both the stiffness coefficients and
the boundary-induced RHS.

\paragraph{Application settings.}
We keep the 3D surfaces fixed and edit the boundaries of their planar texture
domains. A common initial map is constructed by a positive uniform Tutte
embedding followed by 40 SLIM symmetric-Dirichlet iterations with
boundary-constraint weight $10^8$.
The determinant-one tensors induced by this map initialize the radial
$\tanh$ parameterization above. This scene initialization is excluded from
committed-update latency. For the timed edit, it applies an outward radial Gaussian
brush to every boundary: angular standard deviation $0.35$ radians and peak
displacement $0.03$ times the chart's maximum coordinate extent. The edited
boundaries are rescaled so their summed enclosed area matches the total
source-surface area. Every backend starts from the same saved tensor state
and executes five VAdam steps with learning rate $0.001$ and moment factors
$0.9$ and $0.999$, followed by a final evaluation.\\

\textbf{Remark} There is a publicly available CUDA-accelerated version of QHM provided in  the TQHM repository \citep{wang2026tqhm}. It adopts a cuDSS-based linear solver and is not applicable to large-scale meshes with more than 5M vertices due to the time and memory constraints. 
Regarding the comparisons of the  end-to-end performance of the QHM pipelines shown in Table 3 of the main paper, our customized implementation still outperforms TQHM by  $1.9\times$ on the reference example \texttt{lucy\_o\_G} with 48K vertices.
\subsection{Quasi-Harmonic Weights}

\begin{figure*}[h]
  \centering
  \includegraphics[width=\linewidth]{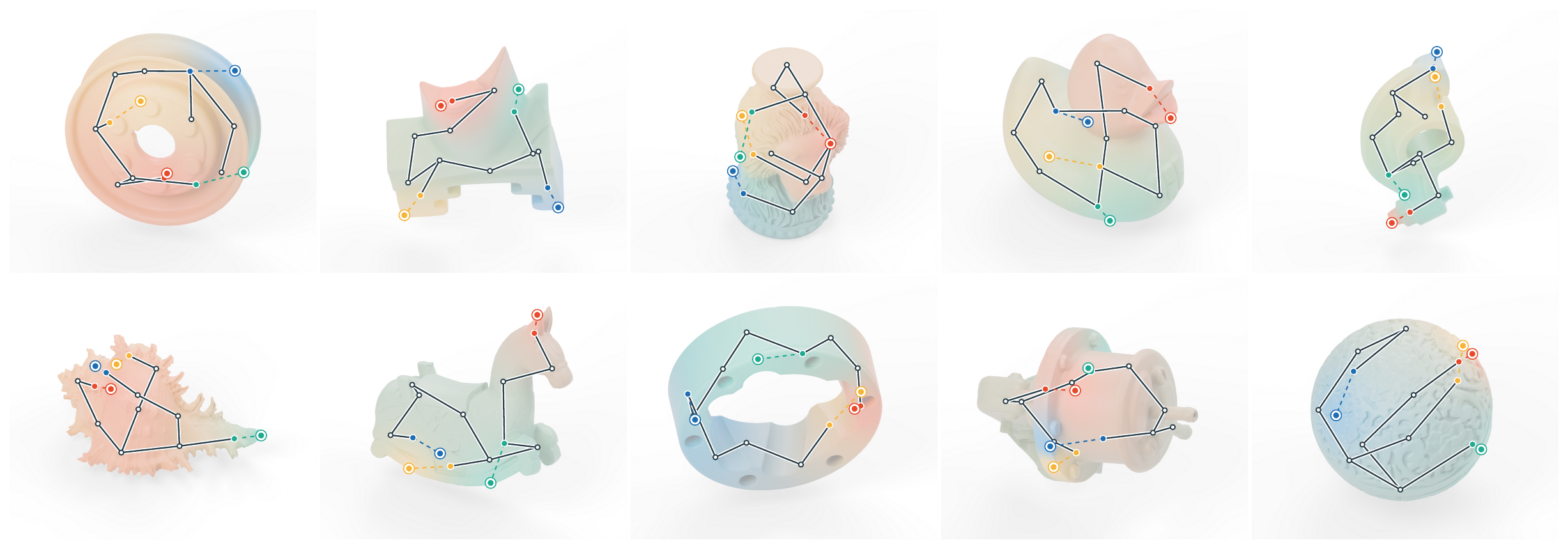}
  \caption{The skinning skeletons for the QHW benchmarking dataset generation.}
  \label{fig:analysis-amgx-block-size}
\end{figure*}

Quasi-harmonic weights (QHW)~\cite{wang2021qhw} optimize diffusion
coefficients on a tetrahedral mesh to obtain interpolation weights.
Four prescribed handles define four primal RHS columns. Differentiation
through the primal solving step requires four adjoint columns sharing
the same constrained matrix.

\paragraph{Diffusion parameterization.}
For the six parameters $\boldsymbol\theta=(\theta_1,\ldots,\theta_6)$
of one tetrahedron, define
\begin{equation}
 \mathbf F(\boldsymbol\theta)=
 \begin{bmatrix}
 1&0&0\\ \theta_2&\theta_3&0\\
 \theta_4&\theta_5&\theta_6
 \end{bmatrix},
 \quad
 \mathbf H=\mathbf F\mathbf F^\top,
 \quad s_i=H_{ii}.
\end{equation}
The implemented diffusion coefficient is
\begin{equation}
 \mathbf C_\tau^{(\nu)}
 =\delta\mathbf I+
 \frac{\mathbf H+\kappa\operatorname{diag}
        (s_2+s_3,s_1+s_3,s_1+s_2)}{\theta_1^2},
\end{equation}
evaluated at the current element parameters.
There is no screening term in the primal equation:
$\mathbf M^{(\nu)}=\mathbf0$.
The prescribed handle values remove the constant mode.
The isotropic diffusion regularizer $\delta\mathbf I$ does not by
itself remove that mode.

\paragraph{Primal and adjoint solving.}
The primal solution $\mathbf W$ has one column per handle, with
one-hot prescribed values. The outer objective penalizes a
biharmonic energy of the projected weights
$\widehat{\mathbf W}=\Pi(\mathbf W)$:
\begin{equation}
 \mathcal E=
 \operatorname{tr}\!\left(
 \widehat{\mathbf W}^{\top}\mathbf K_{\rm ref}
 \mathbf M_{\rm ref}^{-1}\mathbf K_{\rm ref}
 \widehat{\mathbf W}\right).
\end{equation}
Here $\mathbf K_{\rm ref}$ and $\mathbf M_{\rm ref}$ are the fixed
reference stiffness and lumped mass, and $\Pi$ performs differentiable
positive-part projection and row normalization. Adjoint solving uses
the same matrix with zero prescribed values. Its RHS depends on the
primal solution, so the eight columns are not all available at once.

\paragraph{Dataset settings.}
The dataset contains an initial matrix followed by six
optimization stages of 50 accepted updates. It therefore gives 301 matrices and resulting in 2,408 linear systems per mesh. The settings are
$\kappa=0.2$, $\delta=10^{-4}$, initial learning rate $0.01$, and
learning-rate decay $0.5$ between stages.
The parameters satisfy $0.1\leq\theta_1\leq10$ and
$|\theta_i|\leq10$ for $i=2,\ldots,6$.
The handles and reference operators remain fixed, which are generated by assigning a skinning skeleton for each mesh, mimicking the workload of the skinning weight computation.

\paragraph{Application settings.}
The interactive example uses three influences: each object has three corresponding spatial control regions, and their one-hot prescribed values define three shared weight columns over the joint volume. Control centers are initialized by farthest-point selection from native vertices, starting at the minimum-$x$ vertex in each source mesh. The sphere radii
are fixed and computed as $0.6\min(0.075d,0.45s)$, where $d$ is that object's bounding-box diagonal and $s$ is the smallest initial inter-control distance.
For the timed edit, the first control on the first object is displaced by
$(0.18,0.08,0)$ times that object's bounding-box diagonal and snapped to a
native vertex. Its resulting scene-space displacement is replicated for the
corresponding control on every object. All backends replay the same saved
tensor parameters and execute three Adam steps at learning rate $0.002$,
followed by a final evaluation.
\subsection{Directional Shape Total Variation}

Directional shape total variation
(DSTV)~\cite{brokman2024dstv} filters a scalar displacement field
along prescribed surface directions. The benchmark records the
reweighted scalar diffusion systems.

\paragraph{Linear systems.}
Let $\mathbf u^{(\nu)}$ be the scalar field before one diffusion step.
On the fixed input surface, define
\begin{equation}
 w_\tau^{(\nu)}
 =\left(\|\mathbf G_\tau\mathbf u^{(\nu)}\|_2+\varepsilon\right)^{-1}.
\end{equation}
The next field is the solution of
\begin{equation}
 \left[\mathbf M_{\rm ref}
 +\frac{h}{2}\sum_\tau|\tau|w_\tau^{(\nu)}
                  \mathbf G_\tau^\top\mathbf G_\tau\right]
 \mathbf u^{(\nu+1)}
 =\mathbf M_{\rm ref}\mathbf u^{(\nu)}.
\end{equation}
The solution $\mathbf x^{(\nu)}$ of the corresponding benchmark
linear system is therefore $\mathbf u^{(\nu+1)}$.
The main-paper coefficients are
$\mathbf C_\tau^{(\nu)}=(h/2)w_\tau^{(\nu)}\mathbf I$ and
$\mathbf M^{(\nu)}=\mathbf M_{\rm ref}$.
The positive mass and lagged weights yield an SPD matrix.
The additive $\varepsilon=0.05$ and the factor $1/2$ follow the
recorded discretization.

\paragraph{Dataset generation.}
Three base-diffusion steps determine the reference displacement and
its directions. For normalized benchmark coordinates, the base step
size is
$10\,\operatorname{median}_i
((M_{\rm ref})_{ii}/(K_{\rm ref})_{ii})$.
The scalar step size is calibrated below one by dyadic bracketing
and six bisections.  The sequence stops when the increment reaches 1\% of the first increment. The current benchmark contains 3,029 scalar linear systems.
\subsection{Implicit MCF Optimization}

Inverse optimization through implicit mean-curvature flow
(IMCF) seeks a surface whose implicit flow step matches a target
surface, following the inverse geometry-processing setting of
Iskra~\cite{dodik2026iskra}.
Here \emph{inverse} refers to optimization through the flow computation,
not a flow whose normal speed is the reciprocal of mean curvature.

\paragraph{Forward and adjoint linear systems.}
Let $\mathbf X$ contain the current vertex positions and
$\mathbf X_\star$ the target positions.
For flow step size $h>0$, the forward computation solves
\begin{equation}
 [\mathbf M(\mathbf X)+h\mathbf K(\mathbf X)]\mathbf U
       =\mathbf M(\mathbf X)\mathbf X.
\end{equation}
Thus, $\mathbf A^{(\nu)}=\mathbf M(\mathbf X)+h\mathbf K(\mathbf X)$
and the three columns of $\mathbf U$ are coordinate solutions.
The objective is
$\|\mathbf U-\mathbf X_\star\|_{\mathbf M(\mathbf X)}^2$.
Three adjoint columns share $\mathbf A^{(\nu)}$ and use the load
$2\mathbf M(\mathbf X)(\mathbf U-\mathbf X_\star)$.
Differentiation includes the dependence of both mass and stiffness on
the current geometry.

On a fixed reference chart with metric $\mathbf g_\tau(\mathbf X)$,
the stiffness contribution can be expressed using
$\sqrt{\det\mathbf g_\tau}\,\mathbf g_\tau^{-1}$.
The factor $h$ is absorbed into $\mathbf C_\tau^{(\nu)}$ in the
main-paper notation. Mesh connectivity remains fixed while the
geometric coefficients and lumped masses evolve.
A separate fixed operator
$\mathbf M(\mathbf X_\star)+\alpha\mathbf K(\mathbf X_\star)$
converts the Euclidean gradient into an $H^1$ gradient; this auxiliary
solving work is excluded from the evolving-system benchmark.

\paragraph{Dataset generation.}
We use $h\in\{0.001,0.005,0.01,0.02\}$ respectively, 
$\alpha=0.02$, and 30 optimization steps per value of $h$.
Its strong-Wolfe line search uses $c_1=10^{-4}$, $c_2=0.9$,
and initial learning rate $32{,}000$.
Initial evaluations, accepted evaluations, and line-search trials
are saved in their original order, with six linear solving per matrix.
The resulting number of matrices depends on the line search and is
not simply the accepted-step count. Initial guesses are reset between independent groups.

\section{Implementation Details}
\label{supp:implementation}

We expand the implementation details of ParaBMG. The coarsest solver eliminates
inter-block coupling through a fixed elimination tree, whereas the
smoother inverts independent principal blocks. Both prepare dense local
operators when coefficients change and reuse them for all subsequent
V-cycles at that coefficient state.

\subsection{Prepared Coarsest Solve}
\label{sec:method_coarse}
\label{supp:coarsest-solve}

While solving linear system $\nu$, every V-cycle uses the same
SPD coarsest matrix $\bA_L^{(\nu)}$. We use nested-dissection multifrontal
elimination to prepare local inverse and coupling operators during
\emph{update}. Repeated \emph{apply} stages then consist of matrix-vector
products and message gathering along the fixed elimination tree.

\paragraph{Structural preparation.}
During \emph{setup}, nested dissection recursively separates the fixed
coarsest sparsity graph. A \emph{separator} is a set of unknowns whose
removal disconnects the remaining graph into groups with no direct
coupling. Each group's interior unknowns precede the separator in the
ordering, so its interior can be eliminated before the separator is
solved. The elimination tree records these dependencies between groups.
At a tree node $f$, let $I_f$ be the unknowns eliminated
there and $\mathcal B_f$ the remaining unknowns coupled to them in
ancestor nodes, including fill introduced by descendant elimination.
We call $\mathcal B_f$ the ancestor boundary. The sets $I_f$ partition
the coarsest unknowns; the root has an empty ancestor boundary.

The \emph{front} $\mathcal K_f^{(\nu)}$ is the dense local matrix on
which elimination is performed at node $f$. Its rows and columns are
indexed by $I_f\cup\mathcal B_f$: the variables in $I_f$ are eliminated
now, while those in $\mathcal B_f$ remain in the reduced equations for
later elimination at ancestors. The front combines the original coefficients assigned
to this node with the Schur-complement contributions of its children.
Every original coefficient is assigned once; overlapping child
contributions are added through their index maps. These maps, the
elimination tree, and the front dimensions depend on the structural
pattern and remain fixed across coefficient updates. A front is thus
an assembled intermediate system, not generally the principal submatrix
$\bA_L^{(\nu)}[I_f\cup\mathcal B_f,I_f\cup\mathcal B_f]$ of the
original matrix. The ancestor boundary may include unknowns eliminated
above the immediate parent.

\paragraph{A two-child example.}
Consider the ordered unknown groups $I_1=\{1,2\}$,
$I_2=\{3,4\}$, and separator $I_r=\{5\}$, with no direct coupling
between $I_1$ and $I_2$. For this example, write
\begin{equation}
 \bA_L=
 \begin{bmatrix}
  \mathbf A_{11}&\mathbf0&\mathbf C_1^\top\\
  \mathbf0&\mathbf A_{22}&\mathbf C_2^\top\\
  \mathbf C_1&\mathbf C_2&a_{55}
 \end{bmatrix}.
 \label{eq:supp-front-example}
\end{equation}
Each child has boundary $\{5\}$ and a $3\times3$ front. Assigning the
original separator coefficient $a_{55}$ to the root, the two children
contribute $-\mathbf C_i\mathbf A_{ii}^{-1}\mathbf C_i^\top$ to its
single entry. The root front is therefore
$a_{55}-\sum_{i=1}^2\mathbf C_i\mathbf A_{ii}^{-1}\mathbf C_i^\top$,
rather than the original $a_{55}$. This example also explains why
original separator coefficients and RHS entries must not be copied
into every child front. Only the pivot blocks must be SPD; an
individual contribution front need not be SPD as a whole.

\paragraph{Numerical preparation.}
Suppressing the state superscript temporarily, write an assembled front
as
\begin{equation}
 \mathcal K_f=
 \begin{bmatrix}
   \mathbf S_f&\mathbf C_f^\top\\
   \mathbf C_f&\mathbf F_f
 \end{bmatrix},\qquad
 \mathbf C_f=(\mathcal K_f)_{\mathcal B_f I_f}.
 \label{eq:supp-coarse-front}
\end{equation}
The pivot $\mathbf S_f$ is SPD. A leaf-to-root traversal factorizes
$\mathbf S_f=\mathbf L_f\mathbf L_f^\top$ and prepares
\begin{equation}
 \begin{aligned}
  \mathbf M_f^{(\nu)}
    &=(\mathbf S_f^{(\nu)})^{-1}
     =(\mathbf L_f^{(\nu)})^{-\top}
       (\mathbf L_f^{(\nu)})^{-1},\\
  \mathbf W_f^{(\nu)}
    &=\mathbf C_f^{(\nu)}\mathbf M_f^{(\nu)}.
 \end{aligned}
 \label{eq:method_coarse_preparation}
\end{equation}
Here $\mathbf M_f$ is a local inverse, distinct from the mass matrix
used in the benchmark formulations. The coupling $\mathbf C_f$ is
taken from the assembled front before its storage is overwritten by
factorization. The Schur contribution sent to the parent is
\begin{equation}
 \mathbf H_f=\mathbf F_f-\mathbf W_f\mathbf C_f^\top.
 \label{eq:supp-coarse-schur}
\end{equation}
Thus the parent's pivot contains all descendant contributions before
it is factorized. No numerical factor or inverse is reused after its
matrix coefficients change.

\paragraph{Block-elimination identity.}
The preparation follows the exact factorization
\begin{equation}
 \mathcal K_f=
 \begin{bmatrix}\mathbf I&\mathbf0\\\mathbf W_f&\mathbf I\end{bmatrix}
 \begin{bmatrix}\mathbf S_f&\mathbf0\\\mathbf0&\mathbf H_f\end{bmatrix}
 \begin{bmatrix}\mathbf I&\mathbf W_f^\top\\\mathbf0&\mathbf I\end{bmatrix}.
 \label{eq:supp-coarse-block-factorization}
\end{equation}
For a local load $[\mathbf q_I^\top,\mathbf q_B^\top]^\top$,
elimination replaces the boundary load by
$\mathbf q_B-\mathbf W_f\mathbf q_I$. Once the boundary solution is
known, back substitution gives
\begin{equation}
 \mathbf x_I=\mathbf M_f\mathbf q_I-\mathbf W_f^\top\mathbf x_B.
 \label{eq:supp-coarse-recovery}
\end{equation}
The same stored coupling therefore serves both directions of traversal.

\paragraph{Repeated apply stages .}
Algorithm~\ref{alg:coarse_apply} applies these identities to a restricted
RHS $\mathbf b_L$. Each original RHS entry is introduced only at the
node that eliminates its variable. Child messages contain elimination
contributions, rather than additional copies of the original boundary
load. A parent gathers and sums overlapping messages at the corresponding
front indices. The local eliminated load $\mathbf q_{I_f}$ is retained
until the downward traversal.

\begin{algorithm}[t]
\caption{The prepared coarsest solve during the \emph{apply} stage}
\label{alg:coarse_apply}
\begin{algorithmic}[1]
\renewcommand{\algorithmicrequire}{\textbf{Input:}}
\renewcommand{\algorithmicensure}{\textbf{Output:}}
\REQUIRE $\mathbf b_L$, prepared $\{\mathbf M_f^{(\nu)},\mathbf W_f^{(\nu)}\}$,
tree, and index sets $\{I_f,\mathcal B_f\}$
\ENSURE Coarse correction $\mathbf x_L$
\FOR{each tree depth, from leaves to root}
  \FOR{each node $f$ at this depth in parallel}
    \STATE Gather and sum child messages into
    $\mathbf h_{I_f},\mathbf h_{\mathcal B_f}$; use zeros at leaves
    \STATE Store $\mathbf q_{I_f}\gets(\mathbf b_L)_{I_f}+\mathbf h_{I_f}$
    \IF{$f$ has a parent}
      \STATE Send $\mathbf h_{\mathcal B_f}
        -\mathbf W_f^{(\nu)}\mathbf q_{I_f}$ to the parent
    \ENDIF
  \ENDFOR
\ENDFOR
\FOR{each tree depth, from root to leaves}
  \FOR{each node $f$ at this depth in parallel}
    \STATE Gather recovered ancestor values $(\mathbf x_L)_{\mathcal B_f}$
    \STATE $(\mathbf x_L)_{I_f}\gets
      \mathbf M_f^{(\nu)}\mathbf q_{I_f}
      -(\mathbf W_f^{(\nu)})^\top(\mathbf x_L)_{\mathcal B_f}$
  \ENDFOR
\ENDFOR
\RETURN $\mathbf x_L$
\end{algorithmic}
\end{algorithm}

At the root the boundary term vanishes. Applying
\autoref{eq:supp-coarse-block-factorization} recursively proves that
the two traversals return $(\bA_L^{(\nu)})^{-1}\mathbf b_L$ in exact
arithmetic. Nodes at the same depth are independent once the preceding
depth has completed; a parent must wait for every child contribution.
Within a node, dense row products expose additional parallelism.

\paragraph{Ordering and reuse.}
If $\mathbf Q_L$ denotes the coarsest permutation, the internal matrix
is $\widetilde{\bA}_L=\mathbf Q_L\bA_L\mathbf Q_L^\top$ and its RHS is
$\widetilde{\mathbf b}_L=\mathbf Q_L\mathbf b_L$. Recovery uses
$\mathbf x_L=\mathbf Q_L^\top\widetilde{\mathbf x}_L$.
Propagating the ordering through finer aggregate blocks, as described
in the main paper, is a change of coordinates. Consistent permutations
satisfy
$\widetilde{\bP}_l=\mathbf Q_l\bP_l\mathbf Q_{l+1}^\top$, so
$\widetilde{\bP}_l^\top\widetilde{\bA}_l\widetilde{\bP}_l
=\mathbf Q_{l+1}\bA_{l+1}\mathbf Q_{l+1}^\top$.

\paragraph{Storage and cost.}
Write $m_f=|I_f|$ and $b_f=|\mathcal B_f|$.
Packed symmetric storage for $\mathbf M_f$ and rectangular storage for
$\mathbf W_f$ require
\begin{equation}
 \sum_f\left[\frac{m_f(m_f+1)}2+b_fm_f\right]
 \label{eq:supp-coarse-storage}
\end{equation}
scalar values, excluding symbolic maps, factorization workspace, and
RHS buffers. The two traversals perform
$O(\sum_f(m_f^2+b_fm_f))$ arithmetic, in addition to message gathering.
Dense pivot factorization, inverse formation, coupling formation, and
Schur updates cost $O(m_f^3+b_fm_f^2+b_f^2m_f)$ per front.
Separator sizes therefore control both preparation cost and storage;
this is not a dense inverse of the entire coarse matrix.
\subsection{Local Block Inversion and Update Scheduling}
\label{supp:block-inverse}

For a smoothing block $B_v$ at level $l$, define
$\mathbf D_v=\bA_l^{(\nu)}[B_v,B_v]$ and $m_v=|B_v|$.
Every $\mathbf D_v$ is an SPD principal submatrix. Unlike the frontal
pivots in \autoref{supp:coarsest-solve}, these blocks do not incorporate
Schur complements from other smoothing blocks. Their inverses form the
block-diagonal smoother, while $\bE_l^{(\nu)}$ retains cross-block
coupling and the V-cycle supplies the coarse correction.

\paragraph{Coefficient layout and arithmetic.}
The local coefficient block is stored in packed lower-triangular form.
For zero-based local indices $i,j$, its packed address relative to the
block offset is
\begin{equation}
 a(i,j)=\frac{h(h+1)}2+s,\qquad h=\max(i,j),\quad s=\min(i,j).
 \label{eq:supp-packed-address}
\end{equation}
The inversion kernel reconstructs a full symmetric work matrix from
these FP32 coefficients and performs elimination in FP64. The output
is a full row-major FP32 inverse. Packed coefficient storage and full
inverse storage serve different operations and must not be counted as
the same allocation. The kernel also records the reciprocal of each
original diagonal entry before overwriting the work matrix.

\paragraph{In-place elimination.}
The implementation uses one dense work matrix $\mathbf X$ and one
saved pivot column $\mathbf c$, rather than an augmented
$[\mathbf D_v\mid\mathbf I]$ array. Initially
$\mathbf X=\mathbf D_v$. Algorithm~\ref{alg:supp-block-inverse}
describes the sweep; the row updates use the saved column values
before the pivot column is overwritten. After all $m_v$ pivots,
$\mathbf X=\mathbf D_v^{-1}$ in exact arithmetic.

\begin{algorithm}[t]
\caption{In-place Inversion of One Smoothing Block}
\label{alg:supp-block-inverse}
\begin{algorithmic}[1]
\renewcommand{\algorithmicrequire}{\textbf{Input:}}
\renewcommand{\algorithmicensure}{\textbf{Output:}}
\REQUIRE Packed SPD block $\mathbf D_v$ of size $m_v$
\ENSURE Stored symmetric inverse $\widehat{\mathbf D_v^{-1}}$
\STATE Unpack $\mathbf X\gets\mathbf D_v$ into FP64 workspace
\FOR{$p=0,\ldots,m_v-1$}
  \STATE Save $\mathbf c\gets\mathbf X_{:,p}$ and $\alpha\gets1/c_p$
  \STATE Set $\mathbf X_{:,p}\gets\mathbf e_p$
  \STATE Scale $\mathbf X_{p,:}\gets\alpha\mathbf X_{p,:}$
  \FOR{each $i\ne p$}
    \STATE $\mathbf X_{i,:}\gets\mathbf X_{i,:}-c_i\mathbf X_{p,:}$
  \ENDFOR
\ENDFOR
\STATE Store $\widehat{\mathbf D_v^{-1}}\gets
  \operatorname{FP32}((\mathbf X+\mathbf X^\top)/2)$
\RETURN $\widehat{\mathbf D_v^{-1}}$
\end{algorithmic}
\end{algorithm}

SPD blocks admit the no-pivot elimination in exact arithmetic. Averaging
the two computed orientations removes asymmetry introduced by rounding;
it does not by itself guarantee positive definiteness after conversion
to FP32. The independently recomputed outer residual remains the
solution-acceptance criterion.

\paragraph{GPU execution.}
Blocks of capacity at most 16 use a 16-thread tile, with two tiles in
a 32-thread CUDA block. Larger blocks, up to the supported capacity of
32, use a full warp. Each tile owns its work matrix and pivot-column
buffer in shared memory. A lane supplies a matrix column during row
updates, the reciprocal pivot is broadcast within the tile, and warp
synchronization separates dependent steps. This exposes parallelism
across matrix entries and across independent smoothing blocks while
retaining the sequential pivot order within each block.

For large levels with sufficiently wide blocks, the implementation
groups block capacities into 8, 16, 24, and 32 to reduce unused shared
memory. For a tile of
capacity $c$, the workspace contains $c^2+c$ FP64 values. Inversion
performs $O(m_v^3)$ work once per coefficient update, while each
subsequent inverse application performs $O(m_v^2)$ work. The full stored
inverse requires $m_v^2$ FP32 values in addition to the
$m_v(m_v+1)/2$ packed coefficient values.

\paragraph{Dependency order of a numerical update.}
Galerkin reductions consume the current coefficients, not the local
inverse entries. All coarser coefficients must therefore be formed
before any aliased coefficient storage is overwritten. The following
schedule makes that dependency explicit and also fixes the damping
weights throughout the next PCG solve.

\begin{algorithm}[t]
\caption{Numerical Update of the Prepared Hierarchy}
\label{alg:supp-numeric-update}
\begin{algorithmic}[1]
\renewcommand{\algorithmicrequire}{\textbf{Input:}}
\renewcommand{\algorithmicensure}{\textbf{Output:}}
\REQUIRE New $\bA_0^{(\nu)}$, fixed partitions, prolongators,
coefficient-reduction maps, and elimination tree
\ENSURE Numerical operators for coefficient state $\nu$
\STATE Refresh finest-level packed blocks and cross-block coefficients
\FOR{$l=0,\ldots,L-1$}
  \STATE Evaluate prepared Galerkin reductions using level-$l$ coefficients
  \STATE Write level-$(l+1)$ coefficients, using CSR at the coarsest level
\ENDFOR
\FOR{each smoothing level $l$}
  \STATE Invert its independent principal blocks
  \STATE Advance the retained spectral-estimation vector and fix $\omega_l^{(\nu)}$
\ENDFOR
\STATE Assemble coarsest fronts and refresh $\{\mathbf M_f^{(\nu)},\mathbf W_f^{(\nu)}\}$
\RETURN Updated hierarchy ready for all RHS columns of this matrix
\end{algorithmic}
\end{algorithm}

For the stored inverse, block smoothing applies
$\mathbf z_{B_v}=\widehat{\mathbf D_v^{-1}}\mathbf r_{B_v}$.
The iterate and residual identities in the main paper are exact for
an exact local inverse; rounding of the stored inverse introduces the
usual mixed-precision approximation. Reusing its numerical values is
valid only while the coefficient state is unchanged. Reusing the block
partition, packed addresses, and reduction maps is valid across states
as long as the hierarchy and structural pattern remain fixed.

\section{Hyperparameter Tuning for Solvers}
\label{supp:experiments}

We have tuned the solvers for each problem. For our method ParaBMG, we use one consistent hyperparameter setting for all testing. For other methods like AMGX and BoomerAMG, they are specifically tuned for each sequence of linear solving. 

\subsection{ParaBMG Tuning}
\label{supp:parabmg-tuning}

The important parameters affect both convergence and work per
linear system. We examine scalar versus block smoothing, interpolation width,
coarsest-level size, ordering, and spectral-update budget. The relevant
selection objective is measured numerical-update time plus the time
for all required RHS columns; iteration count alone is insufficient.

\paragraph{Main-comparison configuration.}
ParaBMG uses two block-Jacobi pre-smoothing sweeps and two
post-smoothing sweeps at every non-coarsest level of each V-cycle,
including the finest level. Thus the sweep count is two on each side
of the coarse correction. Both surface and volume problems use
aggregate-aligned smoothing blocks. These choices are fixed across
all 40 mesh/problem combinations in the main comparison.

\paragraph{Sensitivity experiment.}
We process the complete Wheel/LDSE and Wheel/QHW sequences,
with three retained repetitions per configuration. The reference uses
aggregate-aligned block-Jacobi smoothing, interpolation width four,
maximum aggregate size 32, two pre- and two post-smoothing sweeps,
and a coarsest limit of 4,096 unknowns.
Initial spectral estimation uses 12 power iterations and subsequent
updates use one retained-vector iteration.

The two trajectories contain 50 and 301 coefficient matrices,
respectively, and include all their RHS columns. The reference uses
geometric distance-two coarsening and cuDSS-derived ordering for the
native coarsest solver. Each arm changes one setting and uses three
retained repetitions. Its objective is the accumulated backend-core
update-plus-all-RHS cost, excluding one-time setup.

\begin{table}[t]
\centering
\begin{tabular}{lrr}
\toprule
Arm & Wheel LDSE & Wheel QHW \\
\midrule
Reference & 1.000 $\pm$ 0.001 & 1.000 $\pm$ 0.000 \\
Width 2 & 1.078 $\pm$ 0.000 & 1.273 $\pm$ 0.000 \\
Width 8 & 1.013 $\pm$ 0.001 & 1.011 $\pm$ 0.001 \\
Terminal 1024 & 0.996 $\pm$ 0.001 & 1.076 $\pm$ 0.002 \\
Terminal 8192 & 1.003 $\pm$ 0.001 & 0.999 $\pm$ 0.000 \\
12 update power steps & 1.063 $\pm$ 0.001 & 1.027 $\pm$ 0.001 \\
100 update power steps & 1.608 $\pm$ 0.002 & 1.241 $\pm$ 0.000 \\
\bottomrule
\end{tabular}

\caption{Recorded ParaBMG sensitivity study. Values are complete
trajectory costs divided by the corresponding reference mean;
uncertainties are sample standard deviations of three retained totals
divided by that same denominator. Reference costs are 1,692.50\,ms
for Wheel/LDSE and 26,225.10\,ms for Wheel/QHW. ``Width'' changes
interpolation support. The reference settings specified here
define this sensitivity study separately from the main comparison.}

\label{tab:supp-parabmg-tuning}
\end{table}

\paragraph{Interpolation.}
Increasing interpolation width from four to eight
changes the reported costs by at most 1.3\%, whereas width two raises
them by 7.8-27.3\%. A smaller interpolation stencil can therefore
reduce construction work without improving the accumulated cost.

\paragraph{Coarsest terminal size.}
Reducing the coarsest limit from 4,096 to 1,024 adds a hierarchy level.
It approximately preserves Wheel/LDSE cost and raises Wheel/QHW cost
by 7.6\%. The 8,192 limit produces the same hierarchies as the reference,
providing a repeatability check.

\paragraph{Spectral-update budget.}
One retained-vector power step gives the lowest complete cost among
the tested 1-, 12-, and 100-step updates. Twelve steps add 2.7--6.3\%
and 100 steps add 24.1-60.8\%. This result concerns sequential
coefficient trajectories. It does not establish that one power step
is sufficient after an arbitrary coefficient jump; the damping
estimate remains a heuristic rather than a certified spectral bound.

The recorded experiment illustrates why fewer iterations need not
mean less solving time: changing interpolation or the smoothing operator
also changes numerical-update and execution costs.
The one-parameter comparisons do not determine the performance of a
configuration that combines all individually favorable choices.
Small differences also require caution: identical-hierarchy arms can
have mean differences exceeding their within-arm deviations. The
rounded deviations in \autoref{tab:supp-parabmg-tuning} describe
the recorded repetitions and do not establish a precise ranking
between nearly equal configurations.

\subsection{Baseline Tuning}
\label{supp:baseline-tuning}

\paragraph{Selection procedure.}
AMGX~\cite{naumov2015amgx} and
BoomerAMG~\cite{henson2002boomer} are selected separately for
each mesh-problem combination.
Candidate evaluation uses an ordered
subset containing the first, middle, and last linear systems for each application stage (such as one edit for LDSE) or optimizer step.
This gives 33 sampled RHS solves for LDSE, 72 for QHW, 30 or 36 for
IMCF, and three for DSTV, per mesh and candidate.

  Earlier exploratory studies evaluated alternative AMGX and BoomerAMG
  configurations, varying the AMG construction, relaxation parameters,
  and smoothing counts. The best configured solver remains to be in a 
  very small sub-class for our problems, namely V-cycle PCG with Jacobi smoothers.
  We therefore conduct a light-weight hyper-parameter tuning stage within this sub-class.

\paragraph{AMGX candidates.}
The aggregation candidates use \texttt{SIZE\_2} and
\texttt{JACOBI\_L1}, with relaxation and symmetric sweep counts
$(0.8,1)$, $(0.6,1)$, and $(0.8,2)$.
The candidate set also contains a classical PMIS configuration with
\texttt{JACOBI\_L1}, relaxation $0.8$, and one sweep on each side.
The selected configurations are recorded below.
We denote these aggregation choices G0, G1, and G2 in the order above,
and the classical PMIS choice G3. G3 uses D2 interpolation. 

\paragraph{BoomerAMG candidates.}
The recorded candidates use PMIS coarsening, extended+i interpolation,
and $\ell_1$-Jacobi relaxation. The strength-threshold/interpolation-width
pairs are $(0.25,4)$, $(0.5,4)$, and $(0.5,2)$.
Each uses one pre- and one post-sweep and one preconditioning cycle.
We denote the three choices B0, B1, and B2 in the order above.

For AMGX, G0 has the lowest sampled cost on all LDSE meshes and G3
on all IMCF meshes. DSTV instead selects G0 on seven meshes, G1 on
Wheel and Blade, and G2 on Murex. BoomerAMG selects B0 throughout
LDSE, QHW, and IMCF; on DSTV, B2 has the lowest cost on Duck, Murex,
Isidore Horse, and Red Circular Box. The measurements also show why
iteration count alone is insufficient for selection. On Wheel/LDSE,
G3 reduces $\bar n$ from G0's 33.61 to 17.82, but increases the total time from
1,814.1 to 1,911.9~ms. On Wheel/DSTV, G1 has more iterations than
G0 (17.67 versus 14.67), yet a slightly lower sampled cost
(90.7 versus 92.7~ms). The latter is a sample minimum, with no
repeat-based estimate of the significance of the difference.

\section{Additional Benchmarking Results}
\label{supp:additional-results}

Tables~\ref{tab:supp-application-details-1}
and~\ref{tab:supp-application-details-2} expand the main comparison
with mean iteration counts, numerical update times, and times per
iteration for each mesh/problem pair.
MAS iteration counts and times use the same Morton-ordered block
configuration as the main table, including
the independent-group initial-guess resets described in
\autoref{supp:benchmarks}.

We use $\bar n_{\rm cg}$ for the mean outer PCG iteration count
over all matrix/RHS pairs for MAS, AMGX, BoomerAMG, and ParaBMG.
For GravoMG-GPU, $\bar n_{\rm V}$ instead counts fixed-point
V-cycles. $\overline T_{\rm update}$ averages the numerical update
time over matrices, and $\overline T_{\rm iter}$ averages the ratio
of solve time to native iteration count over positive-iteration RHS
solves.
All times are in milliseconds.

\begin{table*}[p]
\centering
\caption{Detailed solver results, part 1 of 2: LDSE and QHW.
Each solver reports mean iteration count $\bar n_{\rm cg}$
($\bar n_{\rm V}$ for GravoMG-GPU), mean update time
$\overline T_{\rm update}$, mean time per iteration
$\overline T_{\rm iter}$, and synthetic cost $T$.
All times are in milliseconds; all values are rounded to one decimal
place. Costs use the same records and statistic as main-paper
Table 1, excluding one-time setup
and the application pipeline. Pulley/QHW MAS uses the complete
301-matrix, 2,408-RHS replay.}
\label{tab:supp-application-details-1}

\begin{tabular*}{\textwidth}{@{\extracolsep{\fill}}ll*{10}{r}@{}}
\toprule
Solver & Metric & \shortstack{Wheel} & \shortstack{Blade} & \shortstack{Vase\\Lion} & \shortstack{Duck} & \shortstack{Rocker\\Arm} & \shortstack{Murex} & \shortstack{Isidore\\Horse} & \shortstack{Pulley} & \shortstack{Oil\\Pump} & \shortstack{Red\\Circular\\Box} \\
\midrule
\multicolumn{12}{c}{\textbf{LDSE}} \\
\midrule
\multirow{4}{*}{MAS} & $\bar n_{\rm cg}$ & 99.3 & 88.9 & 101.8 & 112.1 & 90.2 & 83.9 & 90.6 & 90.9 & 91.3 & 103.4 \\
 & $\overline T_{\rm update}$ & 12.3 & 18.2 & 18.1 & 18.9 & 20.5 & 64.2 & 287.7 & 301.4 & 297.2 & 293.4 \\
 & $\overline T_{\rm iter}$ & 2.1 & 3.1 & 3.0 & 3.1 & 3.1 & 13.0 & 26.2 & 28.5 & 27.4 & 28.6 \\
 & $T$ & 219.5 & 293.4 & 321.5 & 371.7 & 295.9 & 1155.6 & 2657.5 & 2891.1 & 2799.6 & 3247.9 \\
\midrule
\multirow{4}{*}{AMGX} & $\bar n_{\rm cg}$ & 22.9 & 21.9 & 24.3 & 25.0 & 22.9 & 22.2 & 24.3 & 23.7 & 23.8 & 26.3 \\
 & $\overline T_{\rm update}$ & 21.6 & 32.2 & 34.1 & 33.2 & 34.6 & 113.2 & 124.3 & 139.4 & 165.8 & 145.2 \\
 & $\overline T_{\rm iter}$ & 1.4 & 2.1 & 2.2 & 2.2 & 2.3 & 11.1 & 13.0 & 15.7 & 15.0 & 15.9 \\
 & $T$ & 52.9 & 78.9 & 87.4 & 88.3 & 86.1 & 358.8 & 439.4 & 511.5 & 524.1 & 562.5 \\
\midrule
\multirow{4}{*}{BoomerAMG} & $\bar n_{\rm cg}$ & 10.3 & 9.1 & 10.3 & 10.7 & 9.7 & 9.5 & 10.4 & 9.5 & 9.9 & 10.9 \\
 & $\overline T_{\rm update}$ & 37.4 & 55.1 & 56.6 & 58.4 & 57.3 & 185.7 & 203.5 & 229.5 & 228.8 & 250.6 \\
 & $\overline T_{\rm iter}$ & 2.0 & 3.0 & 3.1 & 3.1 & 3.1 & 13.8 & 15.8 & 19.2 & 18.4 & 19.3 \\
 & $T$ & 58.3 & 82.5 & 88.0 & 91.6 & 87.4 & 316.6 & 367.8 & 412.7 & 411.3 & 459.8 \\
\midrule
\multirow{4}{*}{GravoMG-GPU} & $\bar n_{\rm V}$ & 13.2 & 8.2 & 10.7 & 12.9 & 9.0 & 8.6 & 11.3 & 7.7 & 12.0 & 10.5 \\
 & $\overline T_{\rm update}$ & 69.2 & 123.3 & 125.4 & 125.8 & 128.7 & 585.8 & 648.0 & 744.0 & 743.4 & 761.3 \\
 & $\overline T_{\rm iter}$ & 2.7 & 5.1 & 5.2 & 5.2 & 5.3 & 33.1 & 37.2 & 43.7 & 43.3 & 44.4 \\
 & $T$ & 104.3 & 164.9 & 181.0 & 192.5 & 176.6 & 869.5 & 1069.2 & 1082.1 & 1261.3 & 1228.9 \\
\midrule
\multirow{4}{*}{ParaBMG} & $\bar n_{\rm cg}$ & 7.1 & 6.3 & 7.4 & 7.6 & 6.9 & 6.8 & 7.3 & 6.7 & 7.2 & 7.5 \\
 & $\overline T_{\rm update}$ & 2.6 & 4.8 & 5.7 & 5.6 & 4.9 & 13.8 & 15.0 & 16.8 & 17.3 & 17.6 \\
 & $\overline T_{\rm iter}$ & 1.0 & 1.5 & 1.6 & 1.6 & 1.6 & 5.4 & 5.9 & 6.5 & 6.7 & 6.7 \\
 & $T$ & 9.5 & 14.3 & 17.2 & 17.7 & 15.9 & 50.6 & 57.8 & 60.2 & 65.2 & 67.9 \\
\midrule
\multicolumn{12}{c}{\textbf{QHW}} \\
\midrule
\multirow{4}{*}{MAS} & $\bar n_{\rm cg}$ & 143.4 & 166.8 & 144.4 & 146.6 & 139.8 & 187.6 & 217.0 & 224.7 & 181.7 & 161.6 \\
 & $\overline T_{\rm update}$ & 12.0 & 18.9 & 18.2 & 18.5 & 20.8 & 63.5 & 283.2 & 299.4 & 301.0 & 302.4 \\
 & $\overline T_{\rm iter}$ & 2.1 & 3.1 & 3.0 & 3.1 & 3.0 & 12.9 & 26.2 & 28.7 & 27.4 & 28.7 \\
 & $T$ & 311.4 & 533.9 & 447.2 & 479.2 & 446.4 & 2490.2 & 5963.2 & 6740.3 & 5270.6 & 4934.8 \\
\midrule
\multirow{4}{*}{AMGX} & $\bar n_{\rm cg}$ & 27.6 & 29.3 & 26.8 & 28.7 & 26.9 & 40.9 & 44.7 & 35.1 & 34.0 & 33.3 \\
 & $\overline T_{\rm update}$ & 21.0 & 32.9 & 33.0 & 35.2 & 42.6 & 113.2 & 123.8 & 138.6 & 138.7 & 143.7 \\
 & $\overline T_{\rm iter}$ & 1.4 & 2.1 & 2.2 & 2.2 & 2.2 & 11.0 & 12.9 & 15.6 & 14.9 & 15.8 \\
 & $T$ & 58.8 & 95.5 & 91.6 & 98.8 & 103.0 & 563.3 & 700.4 & 685.2 & 646.7 & 669.0 \\
\midrule
\multirow{4}{*}{BoomerAMG} & $\bar n_{\rm cg}$ & 12.1 & 9.8 & 10.2 & 11.4 & 10.1 & 10.3 & 13.4 & 11.0 & 10.7 & 10.7 \\
 & $\overline T_{\rm update}$ & 38.5 & 56.3 & 57.4 & 59.2 & 57.7 & 233.8 & 260.7 & 290.4 & 312.6 & 333.5 \\
 & $\overline T_{\rm iter}$ & 2.1 & 3.1 & 3.1 & 3.2 & 3.1 & 14.0 & 15.7 & 19.0 & 18.4 & 19.4 \\
 & $T$ & 63.3 & 86.4 & 89.4 & 95.1 & 89.6 & 379.0 & 471.1 & 499.7 & 509.8 & 541.6 \\
\midrule
\multirow{4}{*}{GravoMG-GPU} & $\bar n_{\rm V}$ & 14.4 & 9.6 & 15.9 & 19.2 & 11.0 & 11.3 & 33.7 & 9.4 & 11.7 & 16.1 \\
 & $\overline T_{\rm update}$ & 66.6 & 117.2 & 120.2 & 120.3 & 124.9 & 563.8 & 623.5 & 718.3 & 718.3 & 736.2 \\
 & $\overline T_{\rm iter}$ & 2.6 & 5.0 & 5.2 & 5.2 & 5.4 & 32.7 & 36.8 & 43.6 & 43.1 & 44.3 \\
 & $T$ & 104.6 & 165.1 & 202.6 & 219.8 & 184.9 & 932.7 & 1861.5 & 1128.3 & 1221.3 & 1450.4 \\
\midrule
\multirow{4}{*}{ParaBMG} & $\bar n_{\rm cg}$ & 7.6 & 6.8 & 7.7 & 7.8 & 7.3 & 8.5 & 8.7 & 8.2 & 8.6 & 8.4 \\
 & $\overline T_{\rm update}$ & 2.6 & 4.7 & 5.6 & 5.6 & 4.9 & 13.8 & 14.9 & 16.5 & 16.8 & 17.3 \\
 & $\overline T_{\rm iter}$ & 1.0 & 1.5 & 1.6 & 1.6 & 1.6 & 5.5 & 5.8 & 6.5 & 6.6 & 6.7 \\
 & $T$ & 9.9 & 15.1 & 17.6 & 17.9 & 16.7 & 59.8 & 65.5 & 70.0 & 73.1 & 74.0 \\
\bottomrule
\end{tabular*}

\end{table*}

\begin{table*}[p]
\centering
\caption{Detailed solver results, part 2 of 2: IMCF and DSTV,
with the same four metric rows per solver and definitions as
\autoref{tab:supp-application-details-1}. All times are in
milliseconds; all values are rounded to one decimal place.}
\label{tab:supp-application-details-2}

\begin{tabular*}{\textwidth}{@{\extracolsep{\fill}}ll*{10}{r}@{}}
\toprule
Solver & Metric & \shortstack{Wheel} & \shortstack{Blade} & \shortstack{Vase\\Lion} & \shortstack{Duck} & \shortstack{Rocker\\Arm} & \shortstack{Murex} & \shortstack{Isidore\\Horse} & \shortstack{Pulley} & \shortstack{Oil\\Pump} & \shortstack{Red\\Circular\\Box} \\
\midrule
\multicolumn{12}{c}{\textbf{IMCF}} \\
\midrule
\multirow{4}{*}{MAS} & $\bar n_{\rm cg}$ & 115.4 & 101.1 & 84.4 & 74.3 & 96.8 & 221.0 & 123.9 & 105.0 & 165.1 & 82.7 \\
 & $\overline T_{\rm update}$ & 4.6 & 3.4 & 3.6 & 3.8 & 4.5 & 15.1 & 11.8 & 7.6 & 7.3 & 7.6 \\
 & $\overline T_{\rm iter}$ & 1.1 & 0.8 & 0.8 & 0.9 & 1.0 & 3.7 & 3.4 & 1.7 & 1.5 & 2.0 \\
 & $T$ & 135.4 & 85.1 & 69.0 & 69.9 & 104.1 & 833.6 & 431.2 & 183.9 & 254.0 & 169.0 \\
\midrule
\multirow{4}{*}{AMGX} & $\bar n_{\rm cg}$ & 12.1 & 11.5 & 12.3 & 10.9 & 11.2 & 11.2 & 12.0 & 10.4 & 13.5 & 11.1 \\
 & $\overline T_{\rm update}$ & 14.1 & 11.0 & 12.1 & 12.4 & 13.0 & 40.4 & 29.8 & 20.5 & 19.0 & 22.6 \\
 & $\overline T_{\rm iter}$ & 0.7 & 0.5 & 0.6 & 0.6 & 0.6 & 1.7 & 1.2 & 0.9 & 0.9 & 1.0 \\
 & $T$ & 22.5 & 17.0 & 19.4 & 19.0 & 20.3 & 59.3 & 44.3 & 29.8 & 30.6 & 33.7 \\
\midrule
\multirow{4}{*}{BoomerAMG} & $\bar n_{\rm cg}$ & 11.6 & 10.4 & 11.0 & 9.7 & 10.1 & 10.1 & 10.5 & 9.3 & 12.3 & 10.1 \\
 & $\overline T_{\rm update}$ & 16.9 & 15.0 & 16.0 & 15.9 & 16.4 & 41.6 & 31.4 & 21.7 & 21.7 & 24.4 \\
 & $\overline T_{\rm iter}$ & 1.0 & 0.8 & 0.9 & 0.9 & 0.9 & 2.3 & 1.8 & 1.3 & 1.2 & 1.4 \\
 & $T$ & 28.1 & 23.6 & 25.7 & 24.7 & 25.9 & 65.2 & 50.2 & 33.4 & 36.8 & 38.2 \\
\midrule
\multirow{4}{*}{GravoMG-GPU} & $\bar n_{\rm V}$ & 38.5 & 7.7 & 12.7 & 7.5 & 7.9 & 9.1 & 12.6 & 5.3 & 32.5 & 13.0 \\
 & $\overline T_{\rm update}$ & 12.5 & 10.9 & 10.3 & 12.0 & 7.0 & 27.9 & 24.9 & 21.9 & 18.2 & 17.0 \\
 & $\overline T_{\rm iter}$ & 12.2 & 11.5 & 10.2 & 12.7 & 2.4 & 12.6 & 18.7 & 19.4 & 13.8 & 12.5 \\
 & $T$ & 483.5 & 99.3 & 140.1 & 108.0 & 25.9 & 142.0 & 260.1 & 125.1 & 468.0 & 179.1 \\
\midrule
\multirow{4}{*}{ParaBMG} & $\bar n_{\rm cg}$ & 8.3 & 7.8 & 8.4 & 7.9 & 7.7 & 8.3 & 8.6 & 6.8 & 12.6 & 6.8 \\
 & $\overline T_{\rm update}$ & 1.2 & 0.8 & 0.9 & 0.9 & 1.1 & 2.5 & 1.8 & 1.1 & 1.0 & 1.3 \\
 & $\overline T_{\rm iter}$ & 0.4 & 0.3 & 0.3 & 0.4 & 0.4 & 1.3 & 1.0 & 0.6 & 0.6 & 0.7 \\
 & $T$ & 4.6 & 3.4 & 3.8 & 3.7 & 4.1 & 13.5 & 10.1 & 5.4 & 8.1 & 6.0 \\
\midrule
\multicolumn{12}{c}{\textbf{DSTV}} \\
\midrule
\multirow{4}{*}{MAS} & $\bar n_{\rm cg}$ & 24.6 & 15.1 & 15.4 & 10.7 & 14.5 & 13.3 & 11.8 & 11.8 & 14.7 & 15.1 \\
 & $\overline T_{\rm update}$ & 4.6 & 3.3 & 3.6 & 3.8 & 4.5 & 14.9 & 13.2 & 7.4 & 7.2 & 7.6 \\
 & $\overline T_{\rm iter}$ & 1.1 & 0.8 & 0.8 & 0.9 & 1.0 & 3.7 & 3.4 & 1.7 & 1.5 & 2.0 \\
 & $T$ & 32.6 & 15.6 & 15.9 & 13.4 & 19.7 & 64.3 & 53.6 & 27.6 & 29.5 & 37.3 \\
\midrule
\multirow{4}{*}{AMGX} & $\bar n_{\rm cg}$ & 10.0 & 7.6 & 6.3 & 5.2 & 6.2 & 3.9 & 4.6 & 5.3 & 6.0 & 8.3 \\
 & $\overline T_{\rm update}$ & 14.5 & 10.3 & 10.9 & 9.2 & 10.9 & 24.5 & 17.7 & 14.4 & 13.1 & 16.6 \\
 & $\overline T_{\rm iter}$ & 0.7 & 0.6 & 0.6 & 0.7 & 0.7 & 2.6 & 1.2 & 0.9 & 0.9 & 1.0 \\
 & $T$ & 21.5 & 15.0 & 14.9 & 12.5 & 15.2 & 34.7 & 23.4 & 19.2 & 18.3 & 24.6 \\
\midrule
\multirow{4}{*}{BoomerAMG} & $\bar n_{\rm cg}$ & 6.3 & 5.9 & 6.0 & 4.8 & 5.7 & 5.2 & 4.4 & 5.3 & 5.7 & 6.6 \\
 & $\overline T_{\rm update}$ & 16.8 & 10.8 & 11.5 & 13.9 & 12.0 & 33.3 & 31.4 & 16.5 & 15.4 & 19.7 \\
 & $\overline T_{\rm iter}$ & 0.9 & 0.7 & 0.7 & 0.9 & 0.8 & 2.4 & 1.8 & 1.2 & 1.1 & 1.3 \\
 & $T$ & 22.6 & 15.0 & 15.9 & 18.1 & 16.6 & 45.5 & 39.3 & 22.7 & 21.6 & 28.3 \\
\midrule
\multirow{4}{*}{GravoMG-GPU} & $\bar n_{\rm V}$ & 16.8 & 2.7 & 3.6 & 2.7 & 2.6 & 2.7 & 4.5 & 2.3 & 4.6 & 19.4 \\
 & $\overline T_{\rm update}$ & 12.1 & 12.1 & 10.9 & 11.8 & 8.8 & 27.7 & 24.2 & 21.4 & 17.4 & 16.8 \\
 & $\overline T_{\rm iter}$ & 12.2 & 11.6 & 10.3 & 12.7 & 2.4 & 12.6 & 18.7 & 19.5 & 13.8 & 12.5 \\
 & $T$ & 217.7 & 43.5 & 47.7 & 45.8 & 15.2 & 61.2 & 107.7 & 66.1 & 80.9 & 258.9 \\
\midrule
\multirow{4}{*}{ParaBMG} & $\bar n_{\rm cg}$ & 4.6 & 3.2 & 4.2 & 3.2 & 3.1 & 3.0 & 3.5 & 3.1 & 3.1 & 4.9 \\
 & $\overline T_{\rm update}$ & 1.1 & 0.8 & 0.9 & 0.9 & 1.1 & 2.5 & 1.8 & 1.1 & 1.0 & 1.3 \\
 & $\overline T_{\rm iter}$ & 0.4 & 0.4 & 0.4 & 0.4 & 0.4 & 1.4 & 1.0 & 0.7 & 0.6 & 0.7 \\
 & $T$ & 3.1 & 1.9 & 2.4 & 2.1 & 2.4 & 6.7 & 5.3 & 3.1 & 2.8 & 4.8 \\
\bottomrule
\end{tabular*}

\end{table*}

\end{document}